\documentclass[notitlepage,prd,twocolumn,showpacs,preprintnumbers,nofootinbib,superscriptaddress,floatfix,tightenlines]{revtex4-1}
\usepackage{mathrsfs}
\usepackage{amssymb}
\usepackage{amsmath}
\usepackage{graphicx}
\usepackage{amsfonts,amsbsy}
\usepackage{bm}
\usepackage{nicefrac}
\usepackage{xcolor}
\usepackage[breaklinks,colorlinks,urlcolor=blue,citecolor=citcolor,linkcolor=lcolor]{hyperref}

\definecolor{lcolor}{rgb}{0.5,0,0}
\definecolor{citcolor}{rgb}{0,0.3,0.0}
\definecolor{ao(english)}{rgb}{0.0, 0.5, 0.0}
\usepackage{comment}

\newcommand{\qhat}{{\hat{q}}}
\newcommand{\fig}{Fig.~}

\newcommand{\eq}{Eq.~}

\def\qhat {\hat{q}}

\begin{document}

\title{Asymmetric transverse momentum broadening in an inhomogeneous medium}

\preprint{}
\author{Yu Fu}
\affiliation{Key Laboratory of Quark and Lepton Physics (MOE) \& Institute of Particle Physics,Central China Normal University, Wuhan 430079, China}
\author{Jorge Casalderrey-Solana}
\affiliation{Departament de Física Quàntica i Astrofísica \& Institut de Ciències del Cosmos (ICC),
Universitat de Barcelona, Martí i Franquès 1, 08028 Barcelona, Spain}
%\affiliation{Departament de Fisica Quantica i Astrofisica \& Institut de Ciencies del Cosmos (ICC),
%Universitat de Barcelona, Marti i Franques 1, 08028 Barcelona, Spain}
\author{Xin-Nian Wang}
\email{xnwang@lbl.gov, current address$^3$}
\affiliation{Key Laboratory of Quark and Lepton Physics (MOE) \& Institute of Particle Physics,Central China Normal University, Wuhan 430079, China}
\affiliation{Nuclear Science Division, Lawrence Berkeley National Laboratory,
CA 94720, Berkeley, USA}

\begin{abstract}
Gradient jet tomography in high-energy heavy-ion collisions utilizes the asymmetric transverse momentum broadening of a propagating parton in an inhomogeneous medium. Such broadening is studied within a path integral description of the evolution of the Wigner distribution for a propagating parton in medium. Going beyond the eikonal approximation of multiple scattering, the evolution operator in the transverse direction can be expressed as the functional integration over all classical trajectories of a massive particle with the light-cone momentum $\omega$ as its mass. With a dipole approximation of the Wilson line correlation function, evolution with the light-cone time $t$ is determined by the jet transport coefficient $\hat q$ that can vary with space and time. In a uniform medium with a constant $\hat q_0$, the analytical solution to the Wigner distribution becomes a typical drifted Gaussian in both transverse momentum and coordinate with the diffusion width $\sqrt{\hat q_0t}$ and $\sqrt{\hat q_0t^3/3\omega^2}$, respectively. In the case of a simple Gaussian-like transverse inhomogeneity with a spatial width $\sigma$ on top of a uniform medium, the final asymmetrical momentum distribution can be calculated semi-analytically. The transverse asymmetry defined for jet gradient tomography that characterizes the asymmetrical distribution is found to linearly correlate with the initial transverse position of the propagating parton within the domain of the inhomogeneity. It decreases with the parton energy $\omega$,  increases with the propagation time initially and saturates when the diffusion distance is much larger than the size of the inhomogeneity or $t^3\gg 3\omega^2\sigma^2/\hat q_0$. The transverse momentum broadening due to the inhomogeneity also saturates at late time in contrast to the continued increase with time if the drifted diffusion in space is ignored. 
\end{abstract}

\maketitle

\section{Introduction}
A jet is essentially a collection of collimated shower of particles stemming from the fragmentation of energetic partons  in high-energy hadron and nuclear collisions. In high-energy heavy-ion collisions, jets also interact with quark-gluon plasma (QGP), a deconfined and strongly coupled state of matter formed in the collisions ~\cite{PHENIX:2004vcz,BRAHMS:2004adc,PHOBOS:2004zne,STAR:2005gfr}, as they travel through the hot and dense matter. The final jet observables therefore should carry the information of jet-medium interaction and are naturally a useful probe to the fundamental properties of QGP.

When jets propagate through the strongly coupled QGP, the energetic partons undergo multiple scattering with the constituents of the QGP and lose their energy, giving rise to the strong attenuation of the high transverse momentum tails of single inclusive hadron spectra as well as the single inclusive jet spectra. These phenomena are usually referred to as jet quenching \cite{Gyulassy:1990ye,Wang:1992qdg}, which has been observed in experiments at the Relativistic Heavy-ion Collider (RHIC) via suppression of large transverse momentum ($p_T$) hadrons \cite{PHENIX:2001hpc,STAR:2002ggv} and later confirmed at the Large Hadron Collider (LHC) via the dijet and $\gamma/Z$-jet asymmetry~\cite{ATLAS:2010isq,CMS:2011iwn} and the suppression of high  $p_T$ particle~\cite{ALICE:2010yje} and jets \cite{ATLAS:2012tjt,ATLAS:2018gwx}. These experimental data have provided important information about the properties of the QGP through jet tomographic studies.

Central to the jet tomography is the energy loss and transverse momentum broadening of a propagating parton inside QGP. Following the first attempt to calculate the parton energy loss in QGP \cite{Gyulassy:1993hr}, several approaches have been established, including  BDMPS-Z\cite{Baier:1996kr,Baier:1996sk,Baier:1998kq,Zakharov:1996fv,Zakharov:1997uu,Zakharov:1998sv}, GLV\cite{Gyulassy:2000fs,Gyulassy:2000er}, ASW\cite{Wiedemann:2000ez,Wiedemann:2000za,Kovner:2003zj,Salgado:2002cd,Armesto:2004ud}, AMY\cite{Arnold:2001ba,Arnold:2001ms,Arnold:2002ja} and Higher-twist\cite{Guo:2000nz,Wang:2001ifa}. In these approaches, the parton energy loss is dictated by the jet transport coefficient $\hat{q}$, which is defined as the averaged transverse momentum broadening squared per unit length \cite{Baier:1996sk}. It has been extracted from comparisons between model calculations and experimental data on single inclusive hadron spectra at both RHIC and LHC ~\cite{Chen:2010te,JET:2013cls,JETSCAPE:2021ehl}.

The initial jet production positions in high-energy nucleus-nucleus collisions in these model calculations are assumed to follow the number of binary nucleon-nucleon collisions  with the Woods-Sexon nuclear distribution~\cite{Miller:2007ri}. The final hadron spectra are averaged over the initial jet production positions and propagation direction. For non-central nucleus-nucleus collisions, the parton propagation length and energy loss will depend on the azimuthal angle of the initial parton propagation direction relative to the reaction plane. This will lead to the azimuthal anisotropy of the final hadron and jet spectra \cite{Wang:2000fq,Gyulassy:2000gk,Gyulassy:2001kr} which in turn can provide information about the path-length dependence of the parton energy loss and the geometrical properties of the QGP \cite{Betz:2011tu,Zigic:2018ovr,Andres:2019eus,Noronha-Hostler:2016eow,Xu:2015bbz,Shi:2018izg,He:2022evt}. One can further use both longitudinal \cite{Zhang:2007ja,Zhang:2009rn} and transverse jet tomography~\cite{He:2020iow} to localize the initial jet production positions and study the space-time profile of the jet transport coefficient in more detail.

The longitudinal jet tomography utilizes the path-length dependence of the parton energy loss and suppression of the final hadron and jet spectra while the transverse jet tomography relies on the asymmetrical transverse momentum broadening of the propagating parton due to the inhomogeneity of the jet transport coefficient in the transverse plane. The latter is also referred to as the gradient jet tomography. It has been applied to localize the initial transverse production positions of $Z/\gamma$-jet to enhance the effect of the diffusion wake induced by $Z/\gamma$-jets in the final $Z/\gamma$-hadron and jet-hadron correlations~\cite{Yang:2021qtl}.

The principle of gradient jet tomography~\cite{He:2020iow} is based on the asymmetrical transverse momentum broadening for an energetic parton propagating in a medium that is inhomogeneous in the transverse direction as characterized by the finite transverse gradient in the jet transport coefficient $\hat{q}$. One can study the asymmetrical transverse momentum broadening via solving a drift-diffusion Boltzmann equation which describes the diffusion of a jet parton in both transverse momentum and coordinate. The finite transverse gradient of the jet transport coefficient in a nonuniform medium leads to a drift in both the final transverse momentum and coordinate distribution of the jet parton which depends on the propagation length and the initial transverse position in the region of the medium with finite gradient of jet transport coefficient \cite{He:2020iow}. One can therefore use the transverse momentum asymmetry of the final jet particles to localize the initial transverse position of the jet production. This principle of gradient tomography has been verified \cite{He:2020iow} by full event-by-event simulations of $\gamma$-jet propagation within the Linear Boltzmann Transport (LBT) model\cite{Li:2010ts,Wang:2013cia,He:2015pra,Luo:2018pto}. It has also been applied to the study of diffusion wake induced by $\gamma/Z$-jets in high-energy heavy-ion collisions \cite{Yang:2021qtl}.

In this study we will formulate the transverse diffusion of a propagating parton in the path integral approach \cite{Zakharov:1997uu,Zakharov:1998sv,Wiedemann:2000za,Kovner:2003zj} for the evolution of parton distributions which are defined via a Wigner function~\cite{Wigner:1932eb,Hillery:1983ms}. Within a picture of multiple soft scattering off independent scattering centers without color flow, the evolution of the Wigner function can be described by the Green's function in  Quantum Chromodynamics (QCD) in a medium that is nonuniform in the transverse plane. One can express the final parton transverse momentum spectrum in terms of the Green's function and calculate the transverse momentum asymmetry and study its path length and transverse gradient dependence. Both our approach with path integral and the drift-diffusion Boltzmann equation assume the dominance of multiple soft scattering in the medium. Asymmetrical transverse momentum broadening due to a few hard parton scatterings has been studied recently in Refs.~\cite{Sadofyev:2021ohn,Barata:2022krd}.

The remainder of this paper is organized as follows. In Sec.~\ref{The propagation of parton in medium}, we briefly review the Wilson line and the Green's function or evolution operator for a parton traveling through a background medium within the multiple soft scattering picture. We will introduce the Wigner function to describe the phase-space distribution of the parton projectile and employ the dipole approximation to related the Wilson line correlations to the jet transport coefficient. We will derive the final phase-space distribution of the propagating parton within the path integral approach. In Sec.~\ref{Spectra in uniform medium}, we derive the final phase-space distribution of a propagating parton in a uniform medium which will be shown to satisfy the drift-diffusion Boltzmann equation. In Sec.~\ref{transverse momentum asymmetry in nonuniform medium}, we calculate the final transverse momentum spectrum, transverse momentum broadening and the transverse momentum asymmetry of a jet parton propagating in a medium with a simple form of transversely inhomogeneous jet transport coefficient. We will examine their dependence on the initial transverse position, propagation length and the energy. In Sec.~\ref{summarys}, we summarize the result.

\section{The propagation of parton in medium \label{The propagation of parton in medium}}

\begin{figure}[ht]
\centering
\includegraphics[width=0.5\textwidth]{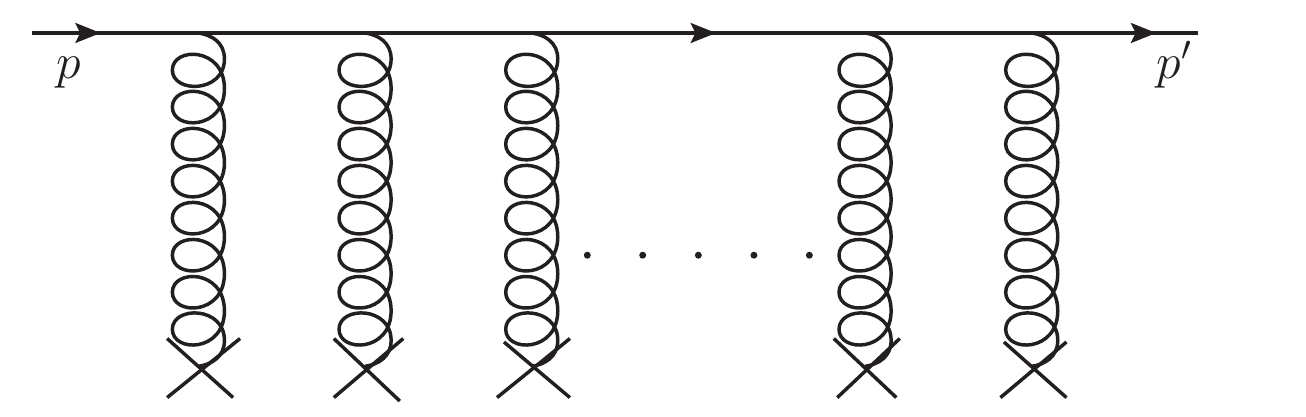}
\caption{Illustration of the eikonal propagation of a quark with initial momentum $p$ and final momentum $p'$ in the medium, where the 'cross' donates medium field.}
\label{fig:multi_scatter}
\end{figure}

To describe the propagation of an energetic parton in a QGP medium within the path-integral approach, we consider multiple soft interaction between a propagating quark and the background field $A(x^+,\bm{x})$ of the QGP medium\footnote{The bold letter refers to the vector lying on the transverse plane throughout this paper.}. Here we adopt the light-cone variables for space-time coordinates,
\begin{equation}
x^{\pm}\equiv\frac{1}{\sqrt{2}}(x^0\pm x^3),
\end{equation}
and similarly for other four-vectors. Assume the initial and final momentum of the quark is $p$ and $p'$, respectively, and $p^+$ is the large component of the momentum, these multiple soft interactions, as illustrated in \fig\ref{fig:multi_scatter}, can be resummed under the eikonal approximation to give the \emph{S-matrix} \cite{Hebecker:1999ej,Wiedemann:2000za,Kovner:2003zj,Casalderrey-Solana:2007knd}
\begin{equation}
S(p',p)\approx 2\pi \delta(p'^{+}-p^{+})2p^{+}\int {\rm{d}}\bm{x}e^{-i(\bm{p}'-\bm{p})\cdot \bm{x}}W(\bm{x}),
 \label{eq:s_eikonal}
\end{equation}
where the Wilson line is defined as
\begin{equation}
W(\bm{x})=\mathcal{P}\exp\Big[ig\int {\rm{d}} x^+ A^-(x^+,\bm{x})\Big],
\label{eq:Wilson_line}
\end{equation}
and $\mathcal{P}$ denotes the path-ordering of field $A^-(x^+,\bm{x})$. 

Under the above eikonal approximation, the sub-leading term $p_{\perp}^2$ in poles of the propagator is ignored. To relax the eikonal approximation, we can keep the $p_{\perp}^2$ term. This is equivalent to considering the Brownian motion of the propagating quark in the transverse plane. In this case, assuming that the initial and final coordinates are $(x^{+}_{i},\bm{x}_{i})$ and $(x^{+}_{f},\bm{x}_{f})$ respectively, the S-matrix can be expressed in the path-integral form \cite{Hebecker:1999ej,Wiedemann:2000za,Kovner:2003zj,Casalderrey-Solana:2007knd},
\begin{widetext}
\begin{equation}
\begin{split}
S(p',p)\approx 2\pi & \delta(p'^{+}-p^{+})2p^{+}\int {\rm{d}}\bm{x} e^{-i(\bm{p'}-\bm{p})\cdot\bm{x}} U(\bm{x}_{f},x^{+}_{f};\bm{x}_{i},x^{+}_{i}),
\end{split}
\end{equation}\\
where
\begin{equation}
\begin{split}
U(\bm{x}_{f},x^{+}_{f};\bm{x}_{i},x^{+}_{i})=\int^{\bm{r}(x^+_f)=\bm{x}_f}_{\bm{r}(x^+_i)=\bm{x}_i}   \mathcal{D}\bm{r}(x^+)\exp[i\frac{p^{+}}{2}\int {\rm{d}}x^+(\frac{d\bm{r} }{dx^+})^2]W(\bm{r}),
\end{split}
\end{equation}
\end{widetext}
is the Green's function or evolution operator\cite{Wiedemann:2000za} that replaces the Wilson line in \eq(\ref{eq:s_eikonal}) and describes the quark's propagation in the transverse plane of the medium. In the adjoint representation, a similar expression for the propagation of a fast gluon with multiple soft scattering has also been derived in Ref.~\cite{Hebecker:1999ej}.

We use the Wigner function \cite{Wigner:1932eb,Hillery:1983ms},
\begin{eqnarray}
\mathcal{W}(\bm{X},\bm{p};x^+)&=& \int {\rm{d}}^{2}\bm{x} e^{-i\bm{p}\cdot\bm{x}} \nonumber \\
&\times&\psi(\bm{X}+\frac{\bm{x}}{2};x^
+)U_{\perp} \psi^{\ast}(\bm{X}-\frac{\bm{x}}{2};x^+),
\end{eqnarray}
to describe the transverse phase-space distribution of a propagating quark at a given time $x^+$,
where $\bm{X}$ and $\bm{p}$ are the transverse coordinate and momentum. The Wilson line $U_\perp$ in this expression makes the Wigner function gauge invariant. It may be taken as a fixed $x^+$ link between the transverse positions of the partonic wave functions. In this work we are interested in the accumulated transverse momentum over a long propagation length in the plasma; therefore, we will neglect its contribution since $U_\perp$ depends on a fixed value of $x^+$. With this approximation and using the evolution operator we can express the evolution of the Wigner function at a later time $x^+_{f}$ from its initial distribution at $x^+_{0}$ with momentum $\bm{p}_0$ as,
\begin{widetext}
\begin{equation}
\begin{split}
\mathcal{W}(\bm{Y},\bm{p};x^+_{f})
=\int {\rm{d}}^{2}\bm{X} {\rm{d}}^{2}\bm{y} {\rm{d}}^{2}\bm{x}\frac{ {\rm{d}}^{2}\bm{p}_0}{(2\pi)^{2}}&e^{-i\bm{p}\cdot(\bm{y}-\bm{x})}\mathcal{W}(\bm{X},\bm{p}_0;x^+_{0})\\
&\Big\langle\!\!\!\Big\langle\! U(\bm{Y}+\frac{\bm{y}}{2},x^+_{f};\bm{X}+\frac{\bm{x}}{2},x^+_{0})
U^{\dagger}(\bm{Y}-\frac{\bm{y}}{2},x^+_{f};\bm{X}-\frac{\bm{x}}{2},x^+_{0})\!\Big\rangle\!\!\!\Big\rangle,
\end{split}
\end{equation}
\begin{equation}
\begin{split}
&\Big\langle\!\!\!\Big\langle\! U(\bm{Y}+\frac{\bm{y}}{2},x^+_{f};\bm{X}+\frac{\bm{x}}{2},x^+_{0})
U^{\dagger}(\bm{Y}-\frac{\bm{y}}{2},x^+_{f};\bm{X}-\frac{\bm{x}}{2},x^+_{0})\!\Big\rangle\!\!\!\Big\rangle\\
=&\int^{\bm{r_1}(x^+_{f})=\bm{Y}+\frac{\bm{y}}{2}}_{\bm{r_1}(x^+_{0})=\bm{X}+\frac{\bm{x}}{2}}\mathcal{D}\bm{r_1} \int^{\bm{r_2}(x^+_{f})=\bm{Y}-\frac{\bm{y}}{2}}_{\bm{r_2}(x^+_{0})=\bm{X}-\frac{\bm{x}}{2}}\mathcal{D}\bm{r_2}
\exp\{i\frac{p^{+}}{2}\int^{x^+_{f}}_{x^+_{0}}\!\!  {\rm{d}} t (\dot{\bm{r_1}}^{2}-\dot{\bm{r_2}}^{2})\}\frac{1}{N_c}\Big\langle\!\!\!\Big\langle tr\{W(\bm{r_1})W^{\dag}(\bm{r_2}) \}\Big\rangle\!\!\!\Big\rangle,
\end{split}
\end{equation}
\\
\end{widetext}
where $\big\langle\!\!\!\!\big\langle\cdots\big\rangle\!\!\!\!\big\rangle$ denotes the average over the proper ensemble of the medium field configurations and $\dot{\bm{r}} \equiv d\bm{r}/d x^+$. Note that the trace and the $1/N_c$ factor correspond to the average over the initial color indices in the fundamental representation.

Under the dipole approximation, the expectation value of the correlation of two Wilson lines can be related to jet transport coefficient $\qhat$, \cite{Zakharov:1997uu,Zakharov:1998sv,Wiedemann:2000ez}
\begin{equation}
\begin{split}
\frac{1}{N_c}\Big\langle\!\!\!\Big\langle tr\{W(\bm{r}_1)W^{\dag}(\bm{r}_2) \}\Big\rangle\!\!\!\Big\rangle &\approx \exp\{-\int^{x^+_{f}}_{x^+_{0}} \!\! {\rm{d}}x^+ \frac{1}{4\sqrt{2}}\hat{q}(\bm{R})\,\bm{r}^2\} \\
&\equiv \exp\{-\int^{t_{f}}_{t_{0}} \!\! {\rm{d}}t \frac{1}{4}\hat{q}(\bm{R})\,\bm{r}^2\} , 
\label{eq:dipole_pprox}
\end{split}
\end{equation}
where $\bm{R}=(\bm{r}_1+\bm{r}_2)/2$ and $\bm{r}=\bm{r}_1-\bm{r}_2$.
In this expression we are implicitly assuming that the transverse separation $\bm{r}\sim 1/{\bm p}$ is much smaller than the scale of variations in $\hat{q}(\bm{R})$.
For convenience, we have re-defined the light-cone variables $t=x^+/\sqrt{2}$  and $\omega=p^+/\sqrt{2}$ which become the normal time and energy on the light-cone.
In a static and homogeneous medium, the correlation will only depend on the relative position of the dipole and the jet transport coefficient will be a constant. Such an approximation is often referred to as the ``harmonic approximation". In this study, however, we generalize the dipole approximation to the case in which the jet transport coefficient has a spatial dependence in the transverse plane. The Wigner distribution of the propagating parton at time $t_f$ is now
\begin{widetext}
\begin{equation}
\begin{split}
\mathcal{W}(\bm{Y},\bm{p};t_f)
=&\int {\rm{d}}^{2}\bm{X} {\rm{d}}^{2}\bm{y} {\rm{d}}^{2}\bm{x}\frac{ {\rm{d}}^{2}\bm{p}_0}{(2\pi)^{2}}e^{-i\bm{p}
\cdot\bm{y}+i\bm{p}_0\cdot\bm{x}}\mathcal{W}(\bm{X},\bm{p};t_{0})\\
&\int^{\bm{R}(t_f)=\bm{Y}}_{\bm{R}(t_0)=\bm{X}}\mathcal{D}\bm{R}\int^{\bm{r}(t_f)=\bm{y}}_{\bm{r}(t_0)=\bm{x}}\mathcal{D}\bm{r}\exp\{i\omega\int^{t_f}_{t_0}\!\!  {\rm{d}}t \dot{\bm{R}}\cdot\dot{\bm{r}}\}\exp\{-\int^{t_f}_{t_0}\!\!  {\rm{d}}t\frac{1}{4}\hat q(\bm{R})\,\bm{r}^2\}.
\label{Wigner1}
\end{split}
\end{equation}
\end{widetext}
To evaluate the path-integral, we discretize the time into $N$ equal steps ($N\to\infty $) and denote $\epsilon=(t_f-t_0)/N$, $t_f=t_N$, $\bm{R}(t_k)=\bm{R}_k$. The second line in Eq.~(\ref{Wigner1}) can be expressed as
\begin{widetext}
\begin{equation}
\begin{split}
K=&\frac{1}{A^{4N}}\Big(\prod_{k=1}^{N-1}\int {\rm{d}}^2\bm{R}_{k}\Big)\exp\Big\{i\omega (\dot{\bm{R}}_N\bm{r}_N-\dot{\bm{R}}_0\bm{r}_0)\Big\}
\Big(\prod_{k=1}^{N-1}\int {\rm{d}}^2\bm{r}_{k}\exp\Big\{-\epsilon\frac{\hat q(\bm{R}_k)}{4}\big(\bm{r}_{k}^2+\frac{4i\omega}{\hat q(\bm{R}_k)}\ddot{\bm{R}}_k\bm{r}_k\big) \Big\}\Big),
\label{K1}
\end{split}
\end{equation}
\end{widetext}
where $A=\sqrt{2\pi i\epsilon/\omega}$. More details on the evaluation of the functional measure can be found in Appendix~\ref{sec:Functional_measure}. Completing the squares and performing Gaussian functional integral to integrate over the relative distance $\bm{r}$, and switching back to the continuous form, Eq.~(\ref{Wigner1}) can be rewritten as
\begin{widetext}
\begin{equation}
\begin{split}
\mathcal{W}(\bm{Y},\bm{p};t_f)=&\int {\rm{d}}^{2}\bm{X} {\rm{d}}^{2}\bm{y} {\rm{d}}^{2}\bm{x}\frac{d^{2}\bm{p}_0}{(2\pi)^{2}}
e^{-i\bm{p}\cdot\bm{y}+i\bm{p}_0\cdot\bm{x}}\mathcal{W}(\bm{X},\bm{p}_0;t_{0})\\
\times&\mathcal{Z}\int^{\bm{R}(t_f)=\bm{Y}}_{\bm{R}(t_0)=\bm{X}} \mathcal{D}\bm{R} \exp\Big\{i\omega \Big(\dot{\bm{R}}(t_f)\bm{y}-\dot{\bm{R}}(t_0)\bm{x}\Big)\Big\}\exp\Big\{-\int^{t_f}_{t_0}\!\!  {\rm{d}} t \frac{\omega^2\ddot{\bm{R}}^2}{\hat q(\bm{R})}\Big\},
\end{split}
\label{Wigner02}
\end{equation}
\end{widetext}
where 
\begin{equation}
    \mathcal{Z}=\frac{2^{3N}\omega^{N}}{4\pi\epsilon^{2N-1}}\frac{1}{\det\{\hat q(\bm{R})\}} 
\end{equation}
is a divergent normalization constant and
\begin{equation}
    \frac{1}{\det\{\hat q(\bm{R})\}}=\prod_{k=1}^{N-1} \frac{1}{\hat q(\bm{R}_k)}.
\end{equation} 
Note that in the above expression, the initial and final transverse momenta are given by the classical momenta of a particle with mass $\omega$ which follows a trajectory $\bm{R}(t)$.

To proceed, we introduce a two-dimensional auxiliary  variable $\bm{\xi}=(\xi^a,\xi^b)$ by defining
\begin{equation}
\bm{\xi}=\omega\ddot{\bm{R}} \quad \text{or} \quad 1=\mathcal{A}\int \mathcal{D}\bm{\xi}\,\delta(\bm{\xi}-\omega\ddot{\bm{R}}),
\end{equation}
where $\mathcal{A}$ is a normalization constant and $\delta(\bm{\xi}-\omega\ddot{\bm{R}})$ is a functional delta-function. We note that $\bm{\xi}$ as so defined can be interpreted as a random force acting on the hard particle with effective mass $\omega$.

Again, after discretizing the space-time, we can write down the expression of $\bm{\xi}$ at $t_k$ as
\begin{equation}	
\bm{\xi}_k=\omega\ddot{\bm{R}}_k=\omega\frac{\bm{R}_{k+1}-2\bm{R}_{k}+\bm{R}_{k-1}}{\epsilon^2}.
\end{equation}
Replacing the variable $\bm{R}$ with $\bm{\xi}$ and using the Jacobian of the transformation,
\begin{widetext}
\begin{equation}
\begin{split}
J\equiv\frac{(\partial\xi^a_1\cdots\partial\xi^a_{N-1})}{(\partial R^a_1\cdots\partial R^a_{N-1})}
=\big(\frac{-2\omega}{\epsilon^2}\big)^{N-1}\!\!
&\times\det
\left( \begin{array}{cccccc}
1 & -\frac{1}{2} & 0 & 0 &\ldots& 0\\
-\frac{1}{2} & 1 & -\frac{1}{2} & 0 &\ldots & 0\\
0 & -\frac{1}{2} & 1 & -\frac{1}{2} &\ldots& 0\\
\vdots & \vdots & \ddots &\ddots &\ddots&\vdots\\
0 & \ldots & 0 & -\frac{1}{2} & 1 & -\frac{1}{2}\\
0 & \ldots & 0 & 0 & -\frac{1}{2} & 1
\end{array} \right)_{\!\!(N-1)\times (N-1)} \;\;\;\;\;\;=N(\frac{\omega}{\epsilon^2})^{N-1},
\end{split}
\end{equation}
we can rewrite Eq.~(\ref{K1}) in terms of a functional integral with respect to $\bm{\xi}$,
\begin{equation}
\begin{split}
K
=&\frac{1}{4\pi^2}\big(\frac{\omega}{t_f-t_0}\big)^2\int\widehat{\mathcal{D}}\bm{\xi}
\exp\Big\{i\omega \Big(\dot{\bm{R}}(t_f)\bm{y}-\dot{\bm{R}}(t_0)\bm{x}\Big)\Big\}\exp\Big\{\!\!-\!\!\int^{t_f}_{t_0}\!  {\rm{d}} t\frac{\bm{\xi}(t)^2}{\hat q(\bm{R})}\Big\},
\end{split}
\end{equation}
and the Wigner function in Eq.~(\ref{Wigner02}) as,
\begin{equation}
\begin{split}
\mathcal{W}(\bm{Y},\bm{p};t_f)
=&\big(\frac{\omega}{t_f-t_0}\big)^2\int {\rm{d}}^{2}\bm{X}  {\rm{d}}^{2}\bm{p}_{0}\mathcal{W}(\bm{X},\bm{p}_{0};t_{0})\\
&\int\widehat{\mathcal{D}}\bm{\xi}\delta^2\big(\omega\dot{\bm{R}}(t_0,\bm{\xi})-\bm{p}_{0}\big)\delta^2\big(\omega\dot{\bm{R}}(t_f,\bm{\xi})-\bm{p}\big)\exp\Big\{-\int^{t_f}_{t_0}  {\rm{d}} t\frac{\bm{\xi}(t)^2}{\hat q(\bm{R})}\Big\},
\label{Wigner2}
\end{split}
\end{equation}
\end{widetext}
where we have defined 
\begin{equation}
\widehat{\mathcal{D}}\bm{\xi}=\frac{1}{\det\{\hat q(\bm{R})\}}\prod_{k=1}^{N-1}\big(\frac{\epsilon}{\pi}\big)d^2\bm{\xi}_{k},
\end{equation}
such that 
\begin{equation}
\int\widehat{\mathcal{D}}\bm{\xi}\exp\Big\{-\int {\rm{d}}t\frac{\bm{\xi}(t)^2}{\hat q(\bm{R})}\Big\}=1.
\end{equation}
See Appendix~\ref{Normalization of Gaussian distribution} for more details on this. With such a normalization condition above, we can interpret the functional integrand 
\begin{equation}
\frac{1}{\det\{\hat q(\bm{R})\}}\exp\Big\{-\int {\rm{d}}t\frac{\bm{\xi}(t)^2}{\hat q(\bm{R})}\Big\},
\end{equation}
as a Gaussian probability distribution of the random force $\bm{\xi}$, driving the Brownian-motion-like  transverse momentum broadening of the propagating parton in the QGP medium. 

With this probability distribution, averaging for any function $f(\bm{\xi})$ over the random variable $\bm{\xi}$ can now be computed as a functional integral,
\begin{equation}
\big\langle f(\bm{\xi})\big\rangle=\int\widehat{\mathcal{D}}\bm{\xi}f(\bm{\xi})\exp\Big\{-\int {\rm{d}}t\frac{\bm{\xi}(t)^2}{\hat q(\bm{R})}\Big\}.
\end{equation}
For parton propagation in a uniform medium with a constant jet transport coefficient $\hat q(\bm{R})=\hat q_0$, the two-point correlation function is (see Appendix~\ref{On correlations} for details),
\begin{eqnarray}
\big\langle\xi^i(t)\xi^j(t')\big\rangle_0&=&\int\widehat{\mathcal{D}}\bm{\xi}\big\{\xi^i(t)\xi^j(t')\big\}\exp\big\{-\int {\rm{d}}t\frac{\bm{\xi}^2}{\hat q_0}\big\} \nonumber \\
&=&\frac{\hat q_0}{2}\delta^{ij}\delta(t-t').
\label{eq:correlator}
\end{eqnarray}

As we noted before, $\bm{\xi}$ can be considered as a random force acting on a particle with an effective mass $\omega$ on a classical trajectory. The boundary conditions for the trajectory are 
\begin{equation}
\bm{R}(t_0,\bm{\xi})=\bm{X},\quad \bm{R}(t_f,\bm{\xi})=\bm{Y},
\end{equation}
at the initial time $t_0$ and the final time $t_f$, respectively, during which the random force $\bm{\xi}$ gives rise to a displacement
\begin{eqnarray}
    \bm{Y}-\bm{X}
    =(t_f-t_0)\dot{\bm{R}}(t_0,\bm{\xi})+\int^{t_f}_{t_0} {\rm{d}} t'\int^{t'}_{t_0} {\rm{d}} t''\frac{\bm{\xi}(t'')}{\omega}. \nonumber\\
\end{eqnarray}
The initial velocity of the particle is therefore
\begin{eqnarray}
\dot{\bm{R}}(&t_0&,\bm{\xi})=\frac{\bm{p}_{0}}{\omega} \nonumber \\
&=&\frac{1}{t_f-t_0}\Big(\bm{Y}-\bm{X}-\int^{t_f}_{t_0} {\rm{d}} t'\int^{t'}_{t_0} {\rm{d}} t''\frac{\bm{\xi}(t'')}{\omega}\Big).
\label{eq::v(t0)}
\end{eqnarray}
The velocity of the particle at any given time $t$ should be
\begin{equation}
\begin{split}
\dot{\bm{R}}(t,\bm{\xi})
&=\dot{\bm{R}}(t_0,\bm{\xi})+\int^{t}_{t_0} {\rm{d}} t''\frac{\bm{\xi}(t'')}{\omega}, 
\end{split}
\label{eq::v(tau)}
\end{equation}
and the position of this particle is given by
\begin{equation}
\bm{R}(t,\bm{\xi})
=\bm{X}+(t-t_0)\frac{\bm{p}_{0}}{\omega}+\frac{1}{\omega}\int^{t}_{t_0} {\rm{d}} t'(t-t')\bm{\xi}(t').
\label{eq:R(tau)}
\end{equation}
In arriving at the last equation for the position $\bm{R}(t,\bm{\xi})$, the following identity is used for an arbitrary smooth function $f(t)$,
\begin{equation}
    \int_{t_0}^t  {\rm{d}} t' \int_{t_0}^{t'}  {\rm{d}} t'' f(t'')=\int_{t_0}^t  {\rm{d}} t'(t-t') f(t').
\end{equation}
Substituting particle's velocities in Eqs.~(\ref{eq::v(t0)})
and (\ref{eq::v(tau)}) into the delta-function in Wigner function in Eq.~(\ref{Wigner2})
, we arrive at
\begin{widetext}
\begin{equation}
\begin{split}
\mathcal{W}(\bm{Y},\bm{p};t_f)=&\big(\frac{\omega}{t_f-t_0}\big)^2\int {\rm{d}}^{2}\bm{X}  {\rm{d}}^{2}\bm{p}_{0}\mathcal{W}(\bm{X},\bm{p}_{0};t_{0})\int\frac{ {\rm{d}}^{2}\bm{x}}{(2\pi)^2}\frac{ {\rm{d}}^{2}\bm{y}}{(2\pi)^2}
\exp\Big\{i\frac{\omega(\bm{Y}-\bm{X})}{t_f-t_0}\cdot(\bm{y}-\bm{x})+i\bm{p}_{0}\cdot\bm{x}-i\bm{p}\cdot\bm{y}\Big\} \\
\times&\int\widehat{\mathcal{D}}\bm{\xi}\exp\Big\{-\int_{t_0}^{t_f}  {\rm{d}} t\frac{\bm{\xi}(t)^2}{\hat q(\bm{R})}\Big\} \exp\Big\{i(\bm{x}-\bm{y})\cdot\int^{t_f}_{t_0} {\rm{d}} t'\int^{t'}_{t_0} {\rm{d}} t''\frac{\bm{\xi}(t'')}{t_f-t_0}+i\bm{y}\cdot\int^{t_f}_{t_0} {\rm{d}} t\bm{\xi}(t)\Big\}
\label{Wigner3}
\end{split}
\end{equation}
Performing the integration over $\bm{p}$ or $\bm{Y}$, we can get the transverse position or transverse momentum distribution of the propagating parton at time $t_f$, respectively,
\begin{equation}
\frac{d^2N}{d^2\bm{Y}}
=(\frac{\omega}{t_f-t_0})^2\!\!\int {\rm{d}}^{2}\bm{X} {\rm{d}}^{2}\bm{p}_{0} \frac{ {\rm{d}}^2\bm{x}}{(2\pi)^2} \mathcal{W}(\bm{X},\bm{p}_{0};t_{0}) e^{-i\bm{x}\cdot(\frac{\omega(\bm{Y}-\bm{X})}{t_f-t_0}-\bm{p}_{0})}\!\! \int\widehat{\mathcal{D}}\bm{\xi}\exp\Big\{i\bm{x}\cdot\int_{t_0}^{t_f}\!\! {\rm{d}} t'\int_{t_0}^{t'}\!\! {\rm{d}} t''\frac{\bm{\xi}(t'')}{t_f-t_0}-\int_{t_0}^{t_f}\!\! {\rm{d}} t\frac{\bm{\xi}(t)^2}{\hat q(\bm{R})}\Big\},
\label{Y_dist}
\end{equation}
%%%%%%
\begin{equation}
\begin{split}
\frac{d^2N}{d^2\bm{p}}
=&\int {\rm{d}}^{2}\bm{X} {\rm{d}}^{2}\bm{p}_{0} \frac{ {\rm{d}}^2\bm{x}}{(2\pi)^2} \mathcal{W}(\bm{X},\bm{p}_{0};t_{0}) e^{-i\bm{x}\cdot(\bm{p}-\bm{p}_{0})}\int\widehat{\mathcal{D}}\bm{\xi}\exp\Big\{i\bm{x}\cdot\int_{t_0}^{t_f} {\rm{d}} t\bm{\xi}(t)-\int_{t_0}^{t_f} {\rm{d}} t\frac{\bm{\xi}(t)^2}{\hat q(\bm{R})}\Big\}.
\label{p_dist}
\end{split}
\end{equation}
\end{widetext}

%%%%%%%%%%%%%%%%%%%%%%%%%%%%%%%%%%%%%%%%%%%%%%%%%%%%%%%%%%%%%%%%%%%%%
\section{Spectra in uniform medium}\label{Spectra in uniform medium}
The general expression for the final phase-space distribution of a propagating parton in terms of path integrals is valid for a inhomogeneous medium in the transverse plane.  In the special case of a uniform QGP medium in which the jet transport coefficient is a constant $\hat{q}(\bm{R})=\hat{q}_0$, one can complete the path integral in Eqs.~(\ref{Wigner3})-(\ref{p_dist}). The distributions can be simplified to a greatest extent as (more details on the derivation are given in Appendix~\ref{sec:Spectra in uniform medium}.)
\begin{widetext}
\begin{equation}
\begin{split}
\mathcal{W}_0(\bm{Y},\bm{p};t_f)=&
\frac{12\omega^2}{\pi^2 {\hat q_0}^2(t_f-t_0)^4}\!\int\!  {\rm{d}}^{2}\bm{X}  {\rm{d}}^{2}\bm{p}_{0}\mathcal{W}(\bm{X},\bm{p}_{0};t_{0})\\
&\exp\Big\{\frac{-12\omega^2}{{\hat q_0}(t_f-t_0)^3}
\Big((\bm{Y}-\bm{X})-\frac{t_f-t_0}{2\omega}(\bm{p}+\bm{p}_{0})\Big)^2-\frac{(\bm{p}-\bm{p}_{0})^2}{{\hat q_0}(t_f-t_0)}\Big\},
\label{eq:general_uni_Wigner}
\end{split}
\end{equation}
\begin{equation}
\begin{split}
\frac{d^2N_0}{d^2\bm{p}}
=&\frac{1}{\pi {\hat q_0}(t_f-t_0)}\int {\rm{d}}^{2}\bm{X}  {\rm{d}}^{2}\bm{p}_{0}\mathcal{W}(\bm{X},\bm{p}_{0};t_{0})\exp\Big\{-\frac{(\bm{p}-\bm{p}_{0})^2}{{\hat q_0}(t_f-t_0)}\Big\},
\label{eq:general_uni_p_dist}
\end{split}
\end{equation}
\begin{equation}
\begin{split}
\frac{d^2N_0}{d^2\bm{Y}}
=\frac{3\omega^2}{\pi {\hat q_0}(t_f-t_0)^3}&\int {\rm{d}}^{2}\bm{X} {\rm{d}}^{2}\bm{p}_{0}\mathcal{W}(\bm{X},\bm{p}_{0};t_{0})
\exp\Big\{-\frac{3(\omega\frac{\bm{Y}-\bm{X}}{t_f-t_0}-\bm{p}_{0})^2}{{\hat q_0}(t_f-t_0)} \Big\},
\label{eq:general_uni_Y_dist}
\end{split}
\end{equation}
\end{widetext}
for any given initial Wigner distribution $\mathcal{W}(\bm{X},\bm{p}_{0};t_{0})$.  

\begin{figure}[ht]
\centering
\includegraphics[width=0.45\textwidth]{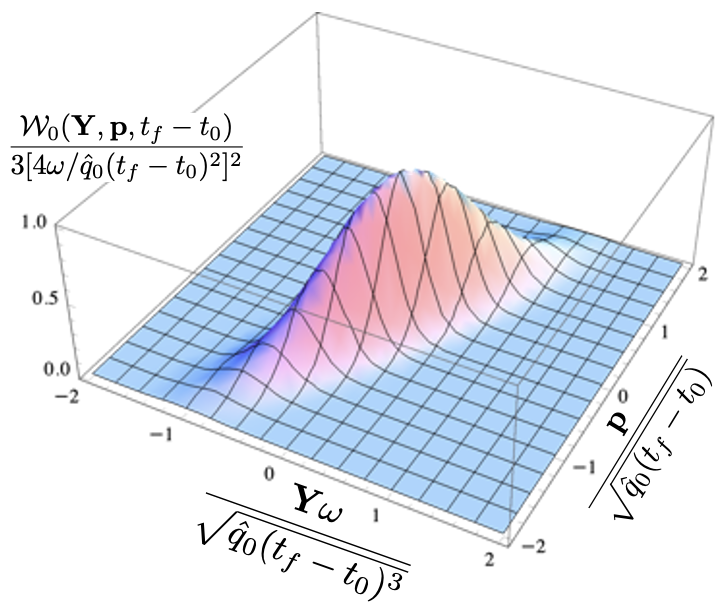}
\caption{The scaled Wigner distribution for an initial classical parton in a uniform medium with a constant jet transport coefficient ${\hat q_0}$ given by Eq.~(\ref{eq:Wigner_constq_point}) as a function of the transverse momentum and coordinate, both scaled by their respective widths, $\sqrt{{\hat q_0}(t_f-t_0)}$ and $\sqrt{{\hat q_0}(t_f-t_0)^3}/\omega$.
\label{fig:w-uniform}}
\end{figure}

For an initial point-like classical particle with specific initial momentum and position $\mathcal{W}(\bm{X},\bm{p}_{0};t_{0})=(2\pi)^2\delta^2(\bm{X})\delta^2(\bm{p}_{0})$, the final Wigner function at a later time $t_f$ becomes
\begin{widetext}
\begin{equation}
\begin{split}
\mathcal{W}_0(\bm{Y},\bm{p};t_f)
&=3\frac{(4\omega)^2}{{\hat q_0}^2(t_f-t_0)^4}
\exp\Big\{\frac{-12\omega^2}{{\hat q_0}(t_f-t_0)^3}(\bm{Y}-\frac{t_f-t_0}{2\omega}\bm{p})^2-\frac{\bm{p}^2}{{\hat q_0}(t_f-t_0)}\Big\}\\
&=3\frac{(4\omega)^2}{{\hat q_0}^2(t_f-t_0)^4}
\exp\Big\{-(\bm{p}-\frac{3\omega}{2(t_f-t_0)}\bm{Y})^2\frac{4}{\hat q_0(t_f-t_0)}-\bm{Y}^2\frac{3\omega^2}{{\hat q_0}(t_f-t_0)^3}\Big\}.
\label{eq:Wigner_constq_point}
\end{split}
\end{equation}
\end{widetext}
One can verify that the final Wigner distribution functions in Eqs.~(\ref{eq:general_uni_Wigner}) and (\ref{eq:Wigner_constq_point}) satisfy the drift-diffusion equation,
\begin{equation}
(\frac{\partial}{\partial t}+\frac{\bm{p}}{\omega}\cdot\bm{\nabla}_{\bm{Y}})\mathcal{W}_0(\bm{Y},\bm{p};t)=\frac{{\hat q_0}}{4}\bm{\nabla}^2_{\bm{p}}\mathcal{W}_0(\bm{Y},\bm{p};t),
\label{eq:dift_diffusion}
\end{equation}
which is just the Boltzmann equation under the special approximation of small-angle scattering whose solution for an initial classical point-like particle in a uniform medium as shown in Eq.~(\ref{eq:Wigner_constq_point}) was first obtained in Ref.~\cite{He:2020iow}. Indeed, as shown in Fig.~\ref{fig:w-uniform}, apart from the usual diffusion in both transverse momentum and coordinate, the Wigner distribution develops a drift $\frac{t_f-t_0}{2\omega}\bm{p}$ in the transverse coordinate for a given value of the transverse momentum $\bm{p}$ and a drift $\frac{3\omega}{2(t_f-t_0)}\bm{Y}$ in the transverse momentum for a given value of the transverse coordinate $\bm{Y}$. The Gaussian diffusion width in the transverse momentum $\sqrt{{\hat q_0}(t_f-t_0)}$ is the typical momentum broadening during the given time interval. The diffusion width in the transverse position is given by the average transverse velocity $\sqrt{{\hat q_0}(t_f-t_0)}/\omega$ times the time interval or $\sqrt{{\hat q_0}(t_f-t_0)^3}/\omega$.

Integrating over the transverse coordinate $\bf{Y}$ or the transverse momentum $\bf{p}$, the Wigner distribution in Eq.~(\ref{eq:Wigner_constq_point}) gives the diffusion distribution in transverse momentum and transverse coordinate, respectively, 
\begin{equation}
\frac{d^2N}{d^2\bm{p}}=\frac{(2\pi)^2}{\pi {\hat q_0}(t_f-t_0)}\exp\Big\{-\frac{\bm{p}^2}{{\hat q_0}(t_f-t_0)}\Big\},
\label{eq:pdist_constq_point}
\end{equation}
\begin{equation}
\frac{d^2N}{d^2\bm{Y}}
=\frac{(2\pi)^2 3\omega^2}{\pi {\hat q_0}(t_f-t_0)^3}\exp\Big\{-\frac{3(\omega\bm{Y})^2}{{\hat q_0}(t_f-t_0)^3} \Big\},
\label{eq:Ydist_constq_point}
\end{equation}
that satisfies the usual Fokker-Planck diffusion equation. We note that the diffusion distribution in the transverse momentum has been obtained within the framework of higher-twist formalism under maximal two-gluon correlation approximation\cite{Liang:2008vz} or by a direct summation of multiple scattering \cite{Majumder:2007hx}.

%%%%%%%%%%%%%%%%%%%%%%%%%%%%%%%%%%%%%%%%%%%%%%%%%%%%%%%%%%%%%%%%
\section{transverse momentum asymmetry in nonuniform medium}\label{transverse momentum asymmetry in nonuniform medium}
To investigate the momentum diffusion in a nonuniform medium within the path integral approach, we consider
a simple transverse distribution of the jet transport coefficient,
\begin{equation}
\hat{q}(\bm{R})=\frac{\hat{q}_0}{1-f(\bm{R})},
\label{eq:qhat}
\end{equation}
with $f(\bm{R})\ll 1$ for all values of $\bm{R}$ which allows us to complete the path integral analytically.

For convenience we denote the functional integral part 
in Eq.~(\ref{p_dist}) for the final momentum spectrum as
\begin{equation}
\begin{split}
F(\bm{x},\bm{X},\bm{p}_{0})=\int\widehat{\mathcal{D}}\bm{\xi}\exp\Big\{\int_{t_0}^{t_f} {\rm{d}} t\Big[
i\bm{x}\cdot\bm{\xi}(t) -\frac{\bm{\xi}(t)^2}{\hat{q}(\bm{R})}\Big]\Big\}.
\end{split}
\label{eq:K_1}
\end{equation}
With the variable transformation $\bm{\xi}'=\bm{\xi}-i\frac{{\hat q_0}}{2}\bm{x}$ and to the leading order in $f$,  it can be approximately rewritten as (see Appendix~\ref{sec:E} for more details)
\begin{equation}
\begin{split}
F(\bm{x},\bm{X},\bm{p}_{0})
\approx &\exp\{ -\int_{t_0}^{t_f}  {\rm{d}} t \frac{{\hat q_0}}{4} \bm{x}^2 \}\\
&\hspace{-1.0cm}\times\Big[1
-\bm{x}^2\int_{t_0}^{t_f}  {\rm{d}} t\frac{{\hat q_0}}{4}
\Big\langle f\big(\bm{R}(t,\bm{\xi}'(t))\big)\Big\rangle_0\Big],
\label{eq:K_2}
\end{split}
\end{equation}
where the average $\Big\langle\cdots\Big\rangle_0$ is defined as
\begin{equation}
\Big\langle\cdots\Big\rangle_0\equiv\int\widehat{\mathcal{D}}\bm{\xi}'\Big(\cdots\Big)\exp\{-\int {\rm{d}}t \frac{\big(\bm{\xi}'(t)\big)^2}{{\hat q_0}}  \}.
\end{equation}

The classical trajectory of the particle presented in Eq.~(\ref{eq:R(tau)}) can be rewritten as
\begin{equation}
\begin{split}
\bm{R}(t,\bm{\xi}')\equiv\bm{R}_0(t)+\Delta\bm{R}(t,\bm{\xi}'),
\end{split}
\end{equation}
with
\begin{equation}
\begin{split}
&\bm{R}_0(t)\equiv\bm{X}+(t-t_0)\frac{\bm{p}_{0}}{\omega}+i\bm{x}\frac{{\hat q_0}}{4\omega}(t-t_0)^2,\\
&\Delta\bm{R}(t,\bm{\xi}')\equiv\frac{1}{\omega}\int^{t}_{t_0} {\rm{d}} t'(t-t')\bm{\xi}'(t').
\end{split}
\end{equation}
Since only $\Delta\bm{R}(t,\bm{\xi}')$ contains the effect of the noise $\bm{\xi}'$, and we can treat it as a perturbation and expand $f(\bm{R})$ perturbatively,
\begin{eqnarray}
f\big(\bm{R}(t,\bm{\xi}')\big)
=&\sum\limits_{n=0}^{\infty}\frac{1}{n!}\!\!\sum\limits_{i_1,\cdots,i_n=1}^{2}\!\!\!\! \Delta R^{i_1}(t,\bm{\xi}')\cdots\Delta R^{i_n}(t,\bm{\xi}') \nonumber \\
&\hspace{-0.7in}\times
\Big(\nabla_{i_1}\nabla_{i_2}\cdots\nabla_{i_n} f\big(\bm{R}(t,\bm{\xi}')\big)\Big)_{\bm{R}(t,\bm{\xi}')=\bm{R}_0(t)},
\label{eq:f_expand}
\end{eqnarray}
where $\nabla_i=\frac{\partial}{\partial R_i}$.
Using the following correlators,
\begin{equation}
\big\langle\Delta R^i(t,\bm{\xi}')\big\rangle_0=0,
\label{eq:R}
\end{equation}
%%%%%
\begin{equation}
\begin{split}
\big\langle\Delta R^i(t,\bm{\xi}')\Delta R^j(t,\bm{\xi}')\big\rangle_0
=&\frac{1}{\omega^2}\frac{{\hat q_0}}{2}\delta^{ij}\frac{(t-t_0)^3}{3},
\label{eq:RR}
\end{split}
\end{equation}
%%%%%
\begin{equation}
\big\langle\Delta R^{i_1}(t,\bm{\xi}')\Delta R^{i_2}(t,\bm{\xi}')\cdots\Delta R^{i_{2n-1}}(t,\bm{\xi}')\big\rangle_0=0,
\label{eq:R_odd}
\end{equation}
\begin{eqnarray}
&\hspace{-1.2in}\big\langle\Delta R^{i_1}\Delta R^{i_2}\!\cdots\!\Delta R^{i_{2n}}(t,\bm{\xi}')\big\rangle_0 \nonumber \\
&=\big\langle\Delta R^{i_1}\Delta R^{i_2}(t,\bm{\xi}')\big\rangle_0
\cdots
\big\langle\Delta R^{i_{2n-1}}\Delta R^{i_{2n}}(t,\bm{\xi}')\big\rangle_0
\nonumber \\
&\hspace{0.2in}+\text{all other permutations},
\label{eq:R_even}
\end{eqnarray}
where total number of permutations is $(2n-1)!!$, we can substitute the expansion of $f(\bm{R})$ in Eq.~(\ref{eq:f_expand}) into Eq.~(\ref{eq:K_2}) and obtain
\begin{widetext}
\begin{equation}
\begin{split}
&F(\bm{x},\bm{X},\bm{p}_{0})\\
=&\exp\{ -\int_{t_0}^{t_f}\!\!  {\rm{d}} t \frac{{\hat q_0}}{4} \bm{x}^2 \} \Big[1- \bm{x}^2\int_{t_0}^{t_f}\!\!  {\rm{d}} t\frac{{\hat q_0}}{4}\sum_{n=0}^{\infty}\frac{2^n(2n-1)!!}{(2n)!}
\big(\frac{{\hat q_0}(t-t_0)^3}{12\omega^2}\big)^{n}  \Big((\bm{\nabla}^2_{\bm{R}})^n f\big(\bm{R}(t,\bm{\xi}')\big)\Big)_{\bm{R}(t,\bm{\xi}')=\bm{R}_0(t)}\Big]\\
=&\exp\{ -\int_{t_0}^{t_f}  {\rm{d}} t \frac{{\hat q_0}}{4} \bm{x}^2 \} \Big[1- \bm{x}^2\int_{t_0}^{t_f}  {\rm{d}} t\frac{{\hat q_0}}{4}\Phi\big(\bm{R}_0(t)\big)\Big],
\end{split}
\end{equation}
\end{widetext}
where 
\begin{equation}
\begin{split}
\Phi\big(\bm{R}(t)\big)&=\sum_{n=0}^{\infty}\frac{(D(t-t_0)^3\bm{\nabla}^2_{\bm{R}})^n}{n!}f(\bm{R}(t)), \\
&\hspace{-0.4in}=\int {\rm{d}}^2\tilde{\bm{R}}\frac{f(\tilde{\bm{R}})}{4\pi D(t-t_0)^3}\exp\left[-\frac{(\bm{R}(t)-\tilde{\bm{R}})^2}{4D(t-t_0)^3}\right],
\end{split}
\label{eq:Phi_1}
\end{equation}
\begin{equation}
\Phi\big(\bm{R}_0(t)\big)=\Phi\big(\bm{R}(t)\big)\Big|_{\bm{R}(t)=\bm{R}_0(t)},
\end{equation}
and $D\equiv {\hat q_0}/12\omega^2$.

With the above approximation of the path integral, we can obtain the transverse momentum distribution, 
\begin{equation}
\frac{d^2 N}{d^2 \bm{p}}\approx \frac{d^2 N_{0}}{d^2 \bm{p}}+\frac{d^2 N_{1}}{d^2 \bm{p}}
\label{eq:N0+N1}
\end{equation}
with $d^2 N_0/d^2 \bm{p}$ given by the solution for a parton propagating in a uniform medium in Eq.~(\ref{eq:general_uni_p_dist}) and
\begin{widetext}
\begin{equation}
\begin{split}
\frac{d^2N_{1}}{d^2\bm{p}}&
=-\int {\rm{d}}^{2}\bm{X}  {\rm{d}}^{2}\bm{p}_{0} \frac{ {\rm{d}}^2\bm{x}}{(2\pi)^2} \mathcal{W}(\bm{X},\bm{p}_{0};t_{0}) 
e^{-i\bm{x}\cdot(\bm{p}-\bm{p}_{0})}\exp\{ -\int_{t_0}^{t_f} {\rm{d}}t \frac{{\hat q_0}}{4} \bm{x}^2 \} \Big[ \bm{x}^2 \int_{t_0}^{t_f} {\rm{d}}t\frac{{\hat q_0}}{4}\Phi\big(\bm{R}_0(t)\big)\Big],
\end{split}
\end{equation}
is the correction linear in $f(\bm{R})$ due the inhomogeneity of the medium.  This linear correction can be rewritten as
\begin{equation}
\begin{split}
\frac{d^2 N_{1}}{d^2 \bm{p}}
=-\int_{t_0}^{t_f} {\rm{d}}t\frac{{\hat q_0}}{4}
\int {\rm{d}}^{2}\bm{X}  {\rm{d}}^{2}\bm{p}_{0}  \mathcal{W}(\bm{X},\bm{p}_{0};t_{0}) G(\bm{p},\bm{p}_{0};\bm{X},t_f, t,t_0),\\
\end{split}
\end{equation}
where the evolution function is defined as
\begin{equation}
\begin{split}
G(\bm{p},\bm{p}_{0};\bm{X},t_f, t,t_0)&=
\int \frac{ {\rm{d}}^2\bm{x}}{(2\pi)^2} 
e^{-i\bm{x}\cdot(\bm{p}-\bm{p}_{0})}\exp\{ -\int_{t_0}^{t_f} {\rm{d}}t \frac{{\hat q_0}}{4} \bm{x}^2 \} \Big[ \bm{x}^2 
\Phi\big(\bm{R}_0(t)\big)\Big].
\end{split}
\label{eq:green}
\end{equation}
\end{widetext}

For a Gaussian form of $f(\bm{R})$,
\begin{equation}
f(\bm{R})=\delta\exp\big\{\frac{-\bm{R}^2}{\sigma^2}\big\},
\label{eq:qhat1}
\end{equation}
the jet transport coefficient in Eq.~(\ref{eq:qhat}) describes a medium that has an increased density within a radius of $\sigma$ in a uniform medium.  One can complete the integration in Eq.~(\ref{eq:Phi_1}) and obtain,
\begin{equation}
\begin{split}
\Phi(\bm{R}(t))
=\frac{\sigma^2\delta}{\Sigma(t)}\exp \Big\{\frac{-\bm{R}(t)^2}{\Sigma(t)}\Big\},
\end{split}
\end{equation}
where
\begin{eqnarray}
    \Sigma(t) \equiv \sigma^2+4D(t-t_0)^3 =\sigma^2+\frac{\hat q_0(t-t_0)^3}{3\omega^2}.
\end{eqnarray}
The evolution function in Eq.~(\ref{eq:green}) becomes,
\begin{widetext}
\begin{equation}
\begin{split}
G(\bm{p},\bm{p}_{0};&\ \bm{X},t_f, t ,t_0)=\int \frac{ {\rm{d}}^2\bm{x}}{(2\pi)^2} 
e^{-i\bm{x}\cdot(\bm{p}-\bm{p}_{0})}\exp\{ -\frac{{\hat q_0}({t_f}-{t_0})}{4} \bm{x}^2 \} \Big[ \bm{x}^2 \frac{\sigma^2\delta}{\Sigma(t)}\exp \Big\{\frac{-[\bm{X}+(t-t_0)\frac{\bm{p}_{0}}{\omega}+i\bm{x}\frac{{\hat q_0}}{4\omega}(t-t_0)^2]^2}{\Sigma(t)}\Big\}
 \Big]\\
=&\frac{1}{(2\pi)^2}\frac{-16\pi\sigma^2\delta}{\Sigma(t)\Delta^3(t)} \exp \Big\{\frac{-\big(\bm{X}+(t-t_0)\frac{\bm{p}_{0}}{\omega} \big)^2}{\Sigma(t)}\Big\} 
\exp\{-\frac{[\bm{p}-\bm{p}_{0}+\lambda(t) \big(\bm{X}+(t-t_0)\frac{\bm{p}_{0}}{\omega}\big)]^2}{\Delta(t)}\}\\
&\times\Big\{[\bm{p}-\bm{p}_{0}+\lambda(t) \big(\bm{X}+(t-t_0)\frac{\bm{p}_{0}}{\omega}\big)]-\Delta(t)\Big\}
\end{split}
\end{equation}
\end{widetext}
where 
\begin{eqnarray}
\Delta(t)&\equiv& {\hat q_0}(t_f-t_0)\big(1-\frac{3}{4}\frac{t-t_0}{t_f-t_0}\frac{4D(t-t_0)^3}{\Sigma(t)}\big) \nonumber \\
&=&{\hat q_0}(t_f-t_0)\big(1-\frac{3}{4}\frac{t-t_0}{t_f-t_0}\frac{\hat q_0 (t-t_0)^3}{3\omega^2\Sigma(t)}\big), \\
\lambda(t)&\equiv&\frac{{\hat q_0}(t-t_0)^2}{2\omega\Sigma(t)}.
\end{eqnarray}
For a parton with initial transverse momentum $\bm{p}_0$ produced at ${\bf x}\equiv(x, y)$ the corresponding initial Wigner function is
\begin{equation}
\begin{split}
\mathcal{W}(\bm{X},\bm{p};t_{0})
&=(2\pi)^2\delta^2({\bf p}-{\bf p}_0)\delta^2({\bf X}-{\bf x})
\end{split}
\end{equation}
The final transverse momentum distribution at time $t_f$ in \eq~(\ref{eq:N0+N1}) is now,
\begin{equation}
\frac{d^2 N_{0}}{d^2\bm{p}}=
\frac{4\pi}{{\hat q_0}(t_f-t_0)}\exp\{-\frac{(\bm{p}-\bm{p}_{0})^2}{{\hat q_0}(t_f-t_0)}\},
\end{equation}
\begin{widetext}
\begin{equation}
\begin{split}
\frac{d^2 N_{1}}{d^2\bm{p}}
=& \int_{t_0}^{t_f}  {\rm{d}} t\frac{4\pi {\hat q_0}\sigma^2\delta }{\Sigma(t)\Delta(t)}\exp\Big\{-\frac{({\bf x}+(t-t_0)\frac{\bm{p}_{0}}{\omega})^2}{\Sigma(t)}-\frac{\big[\bm{p}-\bm{p}_{0}+\lambda(t)({\bf x}+(t-t_0)\frac{\bm{p}_{0}}{\omega})\big]^2}{\Delta(t)} \Big\}\\
&\times\frac{1}{\Delta^2(t)}\Big\{\big[\bm{p}-\bm{p}_{0}+\lambda(t)({\bf x}+(t-t_0)\frac{\bm{p}_{0}}{\omega})\big]^2-\Delta(t)\Big\}.
\end{split}
\end{equation}
\end{widetext}
For a parton with initial transverse momentum ${\bf p}_0=0$, the above final distributions become
\begin{equation}
\frac{d^2N_{0}}{d^2\bm{p}}=
\frac{4\pi}{{\hat q_0}(t_f-t_0)}\exp\{-\frac{\bm{p}^2}{{\hat q_0}(t_f-t_0)}\},
\end{equation}
\begin{equation}
\begin{split}
\frac{d^2 N_{1}}{d^2\bm{p}}
= & \int_{t_0}^{t_f}  {\rm{d}} t\frac{4\pi {\hat q_0}\sigma^2\delta }{\Sigma(t)\Delta^3(t)}\exp\Big\{-\frac{{\bf x}^2}{\Sigma(t)}-\frac{\big[\bm{p}+\lambda(t){\bf x}\big]^2}{\Delta(t)}\Big\} \\
&\times \Big\{\big[\bm{p}+\lambda(t){\bf x}\big]^2-\Delta(t)\Big\}.
\end{split}
\label{eq:dn1p0}
\end{equation}
Since $d^2N_0/d^2{\bf p}$ is the solution to the diffusion equation in a uniform medium, it is symmetric in the transverse plane independent of the initial position ${\bf x}$. The first order correction $d^2 N_1/d^2{\bf p}$ due to the inhomogeneity of the jet transport coefficient $\hat q({\bf x})$ as given by Eqs.~(\ref{eq:qhat}) and (\ref{eq:qhat1}) is asymmetric in the transverse plane for finite values of the parton's initial position ${\bf x}$.

\begin{figure}[ht]
\centering
\includegraphics[width=0.45\textwidth]{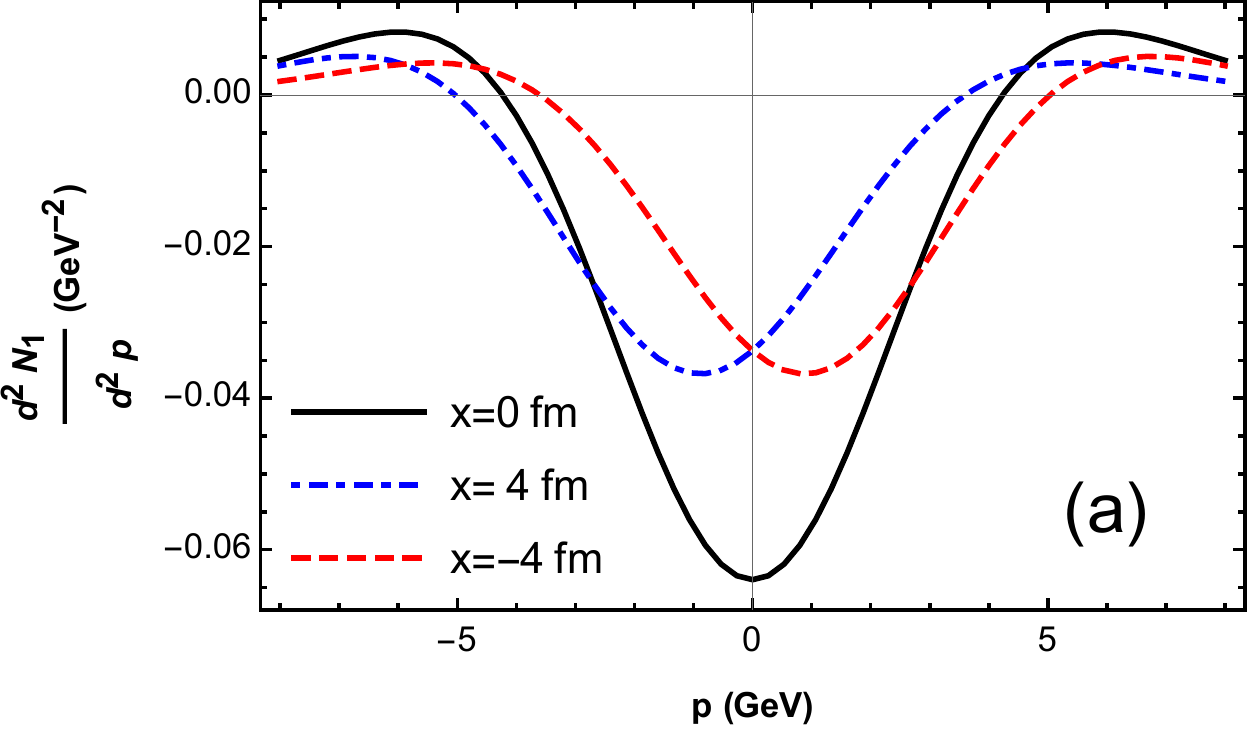}
\includegraphics[width=0.45\textwidth]{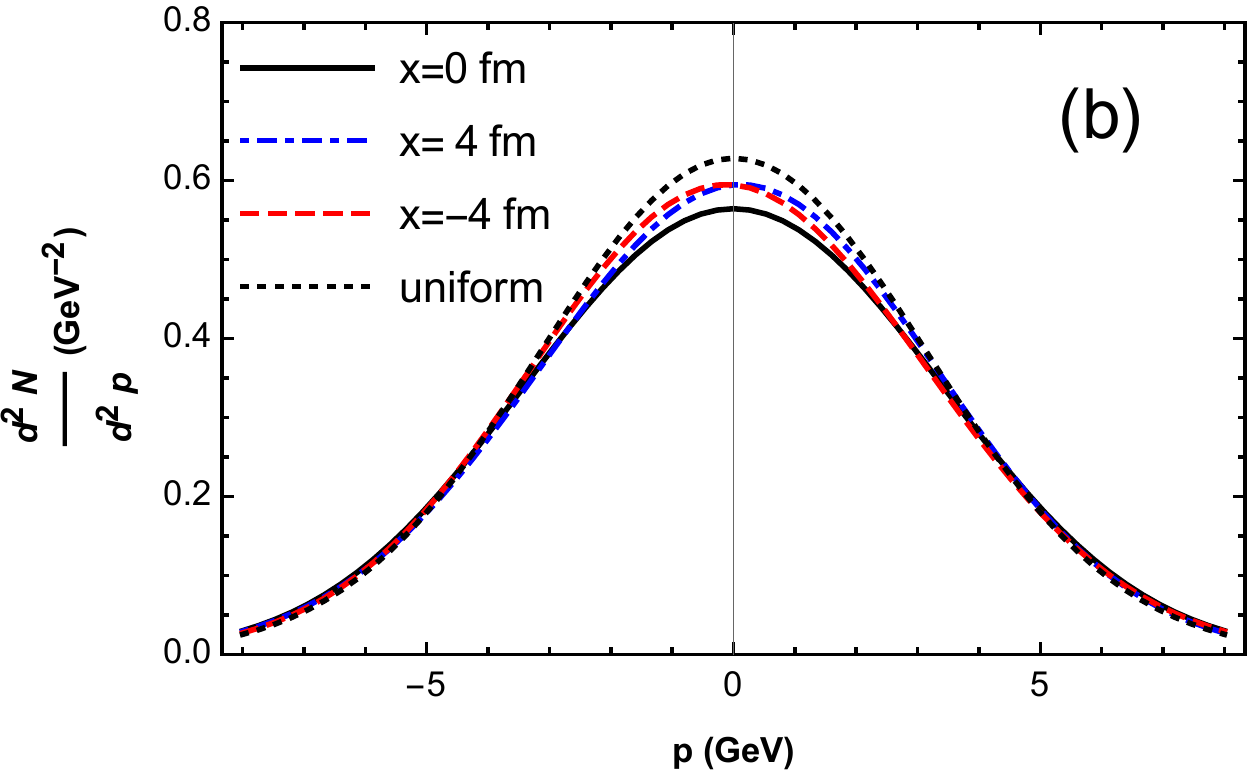}
\caption{(a)The first-order correction and (b) the final transverse momentum distribution for initial position $x=-4.0$ fm (red dashed), 0.0 (black solid), 4.0 (blue dot-dashed) fm,  $\omega=5$ GeV, $t_f-t_0=10$ fm/$c$,  $\hat{q}_0=2$ GeV$^2$/fm,  $\delta=0.1$ and $\sigma=5$ fm in the simple model for an inhomogeneous jet transport coefficient. The dotted line is the distribution in a uniform medium with a constant jet transport coefficient $\hat q_0$.}
\label{fig:distr}
\end{figure}

To illustrate the asymmetrical transverse momentum broadening, we show in Fig.~{\ref{fig:distr}} (a) the first-order correction $d^2N_1/d^2{\bf p}$ and (b) the final transverse momentum distribution as a function of ${\bf p}\cdot \hat{\bf x}$ for ${\bf p}\cdot (\hat {\bf z}\times\hat{\bf x})=0$ and different values of the initial position $|{\bf x}|$. We have set $t_f-t_0=10$ fm/c, $\sigma=5$ fm, $\delta=0.1$, $\hat{q}_0=2$ GeV$^2$/fm, $\omega=5$ GeV and ${\bf p}_0=0$.

In general, the first-order correction in the case we consider here makes the final momentum distribution broader, leading to the increased transverse momentum broadening as compared to that in a uniform medium without a region of inhomogeneity. The distribution is asymmetric for finite values of the initial transverse position of the propagating parton. 

\begin{figure}[ht]
\centering
\includegraphics[width=0.45\textwidth]{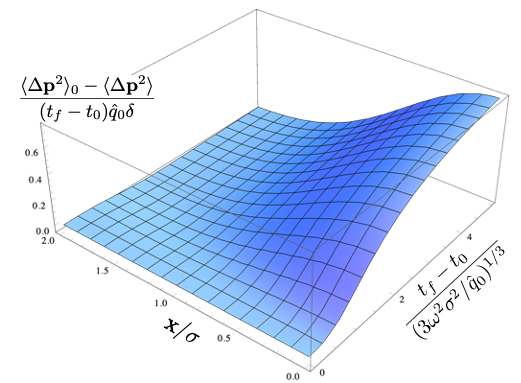}
\caption{The reduction of scaled momentum broadening $(\langle\Delta{\bf p}^2\rangle_0 -\langle\Delta{\bf p}^2\rangle)/(t_f-t_0)\hat q_0\delta$ as a function of the scaled transverse position ${\bf x}/\sigma$ and propagation time $(t_f-t_0)/(3\omega^2\sigma^2/\hat q_0)^{1/3}$.}
\label{fig:broad}
\end{figure}

One can show that the first order correction in Eq.~(\ref{eq:dn1p0}) does not contribute to the zeroth and first moment of the final transverse momentum distribution, $\int {\rm{d}}^2{\bf p} d^2N_1/d^2{\bf p}=\int {\rm{d}}^2{\bf p} {\bf p} d^2N_1/d^2{\bf p}=0$.  However, it increases the total transverse momentum broadening,
\begin{equation}
\begin{split}
    \langle {\bf p}^2\rangle &= (t_f-t_0)\hat q_0+\langle\Delta{\bf p}^2({\bf x},t_f)\rangle, \\
   \langle\Delta{\bf p}^2({\bf x},t_f)\rangle &=\int_{t_0}^{t_f}  {\rm{d}} t\frac{{\hat q_0}\sigma^2\delta }{\Sigma(t) 
   }\exp\Big[-\frac{{\bf x}^2}{\Sigma(t)}\Big],
\end{split}
\label{eq:dpt2}
\end{equation}
due to the extra density of the medium with inhomogeneity in the
region $|x|<\sigma$ on top of a uniform medium. The extra momentum broadening $\langle\Delta{\bf p}^2({\bf x},t_f)\rangle$ due to this region of inhomogeneity grows linearly with the time initially when $t_f-t_0\ll (3\omega^2\sigma^2/\hat q_0)^{1/3}$ and saturates at a finite value asymptotically. At ${\bf x}=0$, this finite extra broadening is $\langle\Delta{\bf p}^2({\bf 0},t_f)\rangle \approx 1.2\hat q_0 \delta (3\omega^2\sigma^2/\hat q_0)^{1/3} $ when $t_f-t_0\gg (3\omega^2\sigma^2/\hat q_0)^{1/3}$. Compared to the extra momentum broadening in this region of transverse inhomogeneity in a scenario of eikonal propagation without the spatial drifted diffusion in the transverse direction,
\begin{equation}
\begin{split}
\langle{\bf p}^2({\bf x},t_f)\rangle_0&=(t_f-t_0)\hat q_0+\langle\Delta{\bf p}^2({\bf x},t_f)\rangle_0 \\
    \langle\Delta{\bf p}^2({\bf x},t_f)\rangle_0
    &\approx (t_f-t_0)\hat q_0 \delta \exp\Big[-\frac{{\bf x}^2}{\sigma^2}\Big],
    \end{split}
    \label{eq:dpt20}
\end{equation}
the drifted diffusion in transverse coordinate due to the transverse gradient reduces the extra momentum broadening in the region of inhomogeneity. Shown in Fig.~\ref{fig:broad} is the reduction of the scaled momentum broadening
$(\langle{\bf p}^2\rangle_0 -\langle{\bf p}^2\rangle)/(t_f-t_0)\hat q_0\delta=(\langle\Delta{\bf p}^2\rangle_0 -\langle\Delta{\bf p}^2\rangle)/(t_f-t_0)\hat q_0\delta$ as a function of the scaled transverse position ${\bf x}/\sigma$ and the scaled propagation time $(t_f-t_0)/(3\omega^2\sigma^2/\hat q_0)^{1/3}$.
One can see that the reduction becomes significant for a propagation time when the transverse drift distance becomes comparable to the size of the inhomogeneity. Since the inhomgeneity-induced broadening in both scenarios dies out exponentially at large $|{\bf x}|>\sigma$ [see Eqs.~(\ref{eq:dpt2}) and (\ref{eq:dpt20})], their difference in Fig.~\ref{fig:broad} also goes to zero exponentially at large $|{\bf x}|$.

The first non-vanishing odd moment of the distribution due to the gradient-induced asymmetrical transverse momentum distribution is
\begin{equation}
\begin{split}
    \langle {\bf p}^3\rangle&=-2{\bf x} \int_{t_0}^{t_f}  {\rm{d}} t\frac{{\hat q_0}\sigma^2\delta }{\Sigma(t) 
   }\lambda(t) \exp\Big\{-\frac{{\bf x}^2}{\Sigma(t)}\Big\} \\
   &=-{\hat q_0}\omega\sigma^2\delta \frac{{\bf x}}{{\bf x}^2}
   \Big[\exp\Big\{-\frac{{\bf x}^2}{\sigma^2+\hat q_0(t_f-t_0)^3/3\omega^2}\Big\} \\
   & \hspace{1.0in} - \exp\Big\{-\frac{{\bf x}^2}{\sigma^2}\Big\} \Big],
   \end{split}
\end{equation}
which grows initially with the cubic of time and saturates at a finite value asymptotically when $(t_f-t_0)^3\gg 3\omega^2\sigma^2/\hat q_0$ because of the finite size of the spatial inhomogeneity.

We can define the transverse asymmetry as proposed in Ref.~\cite{He:2020iow},
\begin{equation}
\begin{split}
A_{N}&=\frac{\int {\rm{d}}^2\bm{Y} {\rm{d}}^2\bm{p}\ \mathcal{W}(\bm{Y},\bm{p},t_f)\ \text{sign}(\hat{\bm{p}}\cdot \hat{\bf x})}{\int {\rm{d}}^2\bm{Y}  {\rm{d}}^2\bm{p}\ \mathcal{W}(\bm{Y},\bm{p},t_f)}\\
&=\int\frac{ {\rm{d}}^2\bm{p}}{(2\pi)^2}\frac{d^2N}{d^2\bm{p}}\ \text{sign}(\hat{\bm{p}}\cdot \hat{\bf x}),
\end{split}
\end{equation}
to characterize the asymmetrical momentum broadening due to the transverse gradient of the medium. Note that the Wigner function is normalized as $\int {\rm{d}}^2\bm{Y} d^2\bm{p}\ \mathcal{W}(\bm{Y},\bm{p},t_f)/(2\pi)^2=1$.
Since the asymmetry is only caused by the first-order correction in Eq.~(\ref{eq:dn1p0}), one can complete the integration over the transverse momentum and obtain the transverse asymmetry as
\begin{equation}
\begin{split}
    A_N=\int_{t_0}^{t_f}  {\rm{d}} t &\frac{{\hat q_0}\sigma^2\delta }{\Sigma(t)\Delta(t)}\frac{\lambda(t)x}{\sqrt{\pi\Delta(t)}}\\
    &\times \exp\Big\{-\frac{{\bf x}^2}{\Sigma(t)}-\frac{\lambda(t)^2{\bf x}^2}{\Delta(t)}\Big\}.
\end{split}
\end{equation}

\begin{figure}[ht]
\centering
\includegraphics[width=0.45\textwidth]{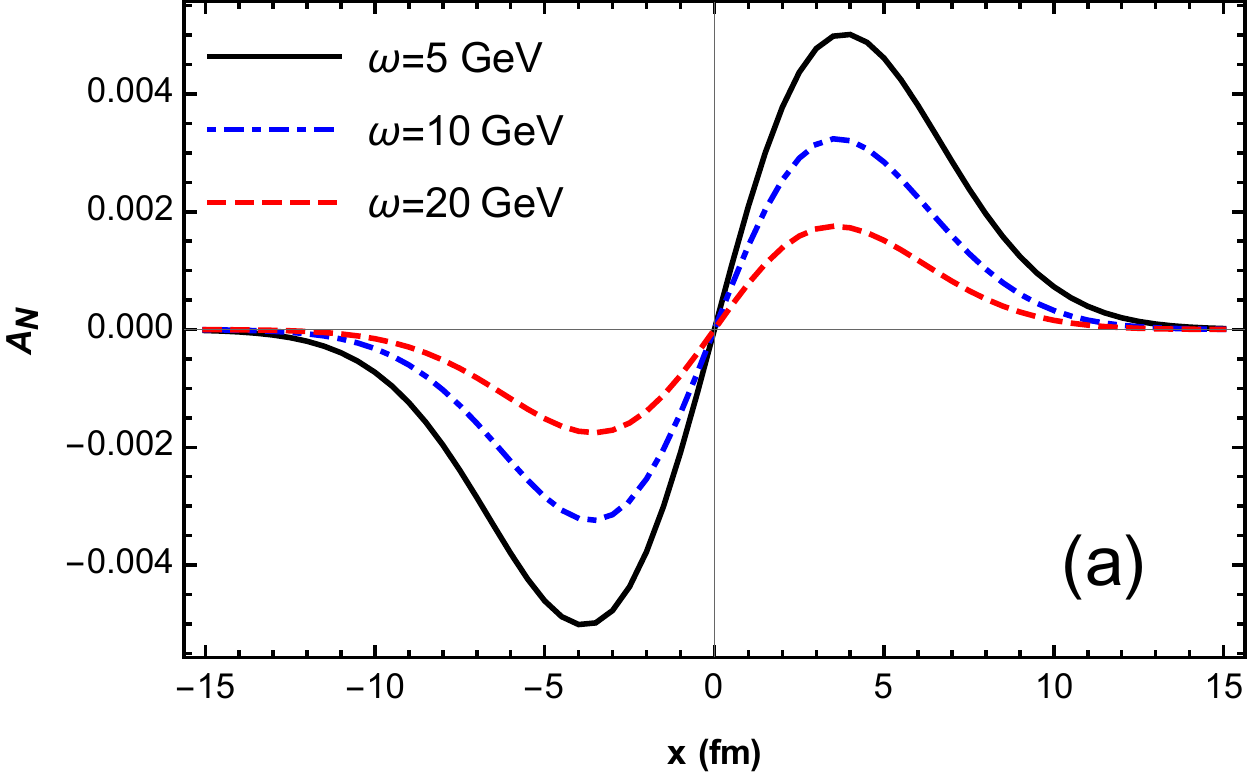}
\includegraphics[width=0.45\textwidth]{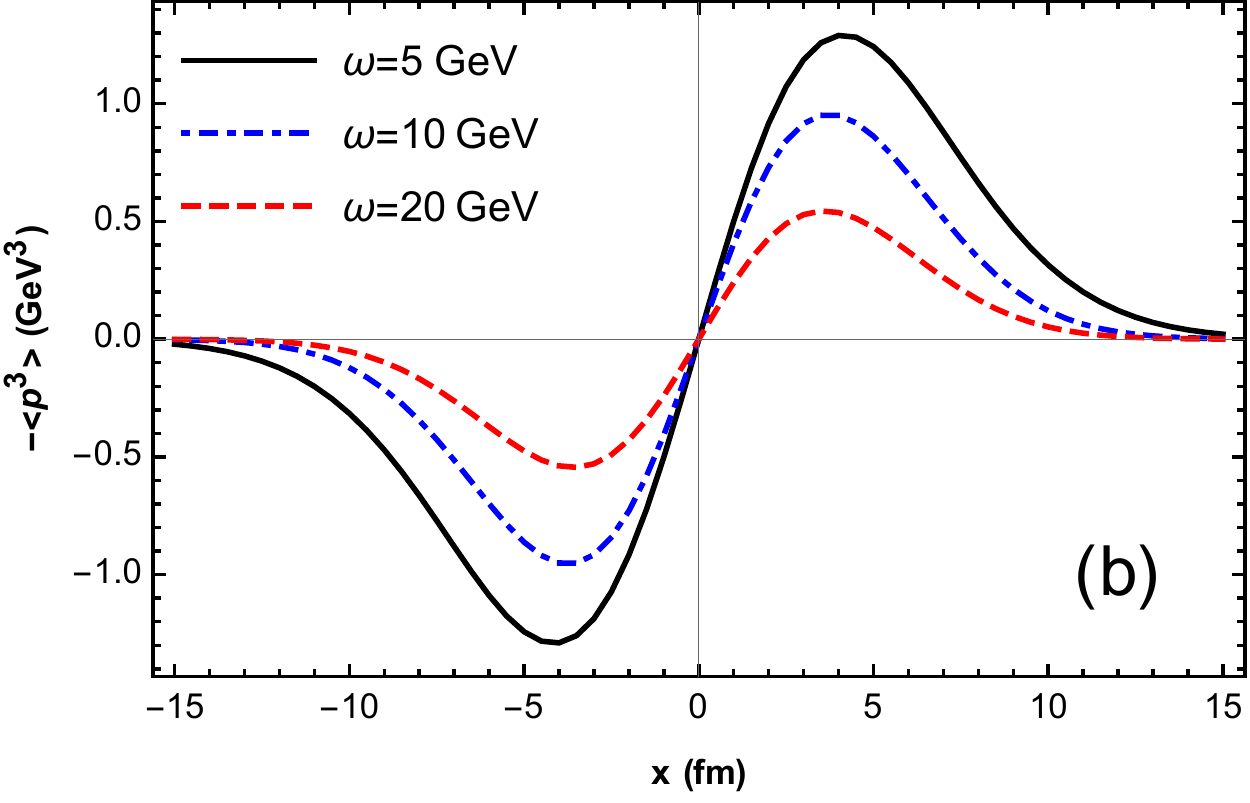}
\caption{Transverse momentum asymmetry as a function of the initial transverse position $ x $ for $t_f-t_0=10$ fm/$c$, $\hat{q}_0=2$ GeV$^2$/fm,  $\delta= 0.1$, $\sigma=5$ fm in the simple model. \label{fig:Asy}}
\end{figure}

The integration over time can be done numerically. Shown in Fig.~\ref{fig:Asy}(a) is the transverse asymmetry $A_N$ as a function of the initial transverse position $x$ for different values of the parton's energy $\omega$. We note that within the size of the transverse inhomogeneity $|x|<\sigma$, the transverse asymmetry is approximately linear in $x$ driven by the transverse gradient. Conversely one therefore can use the transverse asymmetry to infer the initial transverse position of the propagating parton. This is the principle underpinning the gradient jet tomography as proposed in  Ref.~\cite{He:2020iow}. Combined with the longitudinal jet tomography, which uses the longitudinal momentum of the final jet or parton energy loss to constrain the propagation length, the 2D jet tomography can be used to localize the initial jet production position. Outside the range of the medium inhomogeneity $|x|>\sigma$, the transverse asymmetry decreases and vanishes when the transverse gradient diminishes. 

Similar to the second and third moment of the momentum distribution, the transverse asymmetry $A_N$ also increases with the propagation time during the diffusion across the domain of the inhomogeneity. Since $\sqrt{\hat q_0(t_f-t_0)}/\omega$ is the average diffusion velocity, $\sqrt{\hat q_0(t_f-t_0)^3}/\omega$ is the diffusion distance during the propagation time. When this distance is much larger than the size of the inhomogeneity $\sigma$ or $(t_f-t_0)^3\gg \omega^2\sigma^2/\hat q_0$, the transverse asymmetry as well as the increased momentum broadening $\langle\Delta{\bf p}^2\rangle$ and the third moment $\langle{\bf p}^3\rangle$ will saturate to the asymptotic values. Since the average diffusion distance is inversely proportional to the parton's energy $\omega$, the transverse asymmetry $A_N$,  as well as the third moment and the extra momentum broadening, decreases with $\omega$.

In Fig.~\ref{fig:Asy}(b), we also plot the third moment $\langle{\bf p}^3\rangle$ as a function of the initial transverse position $x$. It has the same behavior as the transverse asymmetry. As seen in Fig.~\ref{fig:distr} (a), the first order correction to the distribution changes sign at large transverse momentum. The third moment has a much large weight at large transverse momentum and is therefore dominated by the first order correction in this large momentum region. Therefore, the asymmetry as characterized the third moment has the opposite sign to the transverse asymmetry $A_N$ which is dominated by the distribution at small momentum. However, their dependence on the initial transverse position $x$, the propagation time $t_f-t_0$ and energy $\omega$ is the same.

%%%%%%%%%%%%%%%%%%%%%%%%%%
\section{Summary}\label{summarys}

To demonstrate the principle of the gradient tomography in jet quenching, we have derived the evolution of the Wigner distribution function in transverse momentum and coordinate for a fast parton traveling inside a strong interaction medium within the path integral approach. Within the dipole approximation for the soft multiple scattering in the medium encoded in the correlation of Wilson operators, the evolution can be expressed generally in terms of a Green's function or the evolution operator which is determined by the space-time profile of the jet transport coefficient $\hat q$. 

In a uniform medium with a constant jet transport coefficient $\hat q_0$, one can complete the path integral and obtain the evolution operator and the corresponding Wigner distribution analytically which is also a solution to an drift-diffusion Boltzmann transport equation.  We also considered a special case of inhomogeneous medium by assuming a form of spatial-dependent jet transport coefficient that adds a Gaussian-like region of enhanced medium density with a finite size. The path integral can also be completed in this case and we obtained the evolution operator analytically. We have considered an initial condition for a classical point-like particle and calculated the final transverse momentum distribution and its dependence on the initial transverse coordinate. The distribution is asymmetric when the initial position of the parton is off the center of the Gaussian region because of the transverse gradient. The Gaussian-like inhomogeneity is found to increase the momentum broadening $\langle {\bf p}^2\rangle$ and lead to a nonvanishing value of the odd moment $\langle {\bf p}^3\rangle$ due to the asymmetrical distribution. We also calculated the transverse asymmetry $A_N$ as proposed in the study of the gradient tomography \cite{He:2020iow}. We found both $A_N$ and $\langle {\bf p}^3\rangle$ linearly correlated with the initial transverse position within the region of the inhomogeneity, validating the principle of the gradient jet tomography. This analytical solution also allows us to understand both the propagation time (length) and energy dependence of the transverse asymmetry.

Going beyond the simple form of the spatial-dependent jet transport coefficient, numerical evaluation of the path integral for more realistic case of the medium in high-energy heavy-ion collisions is needed. Since the path integral approach differs from the classical transport approach in which one can also introduce space and time dependence of the jet transport coefficient in the drift-diffusion Boltzmann equation as done in Ref.~\cite{He:2020iow}, it will also be interesting to examine the difference between the two approaches.  These studies will help to establish the gradient jet tomography as a powerful tool to explore properties of QGP using jet quenching.

\begin{acknowledgements}
We thank Yayun He and Longgang Pang for helpful discussions. This work is supported in part by National Natural Science Foundation of China (NSFC) under Grant Nos. 11935007, 11861131009, 11890714, by the Director, Office of Energy Research, Office of High Energy and Nuclear Physics, Division of Nuclear Physics, of the U.S. Department of Energy under  Contract No. DE-AC02-05CH11231, by the US National Science Foundation under Grant No. ACI-1550228 within the JETSCAPE and OAC-2004571 within the X-SCAPE Collaboration and by the grants SGR-2017-754, PID2019-105614GB-C21, PID2019-105614GB-C22 and the ``Unit of Excellence MdM 2020-2023'' award to the Institute of Cosmos Sciences (CEX2019-000918-M).
\end{acknowledgements}

% \#\#\#\#%%%%%%%%%%%%%%%%%%%%%%%%%%%%%%%%%%%%%%%%%%%%%%%%%%%%%%%%%%%
\appendix

\section{Functional measure}\label{sec:Functional_measure}
Let us consider the calculation of the propagator in one-dimensional space. We divide the computation into many steps; inserting the closure relation many times, we write the propagator as the products with small time steps,
\begin{equation}
\begin{split}
\langle x_f&|e^{-iH(t_f-t_0)}|x_0\rangle=\lim_{\epsilon\to0}\langle x_f|e^{-iH\epsilon}e^{-iH\epsilon}\dots e^{-iH\epsilon}|x_0\rangle\\
=&\prod_{i=1}^{N-1}\int {\rm{d}}x_i\langle x_f|e^{-iH\epsilon}|x_{N-1}\rangle\langle x_{N-1}|e^{-iH\epsilon}|x_{N-2}\rangle \\
&\hspace{0.5in}\cdots\langle x_{2}|e^{-iH\epsilon}|x_1\rangle\langle x_1|e^{-iH\epsilon}|x_0\rangle.
\label{eq:app1}
\end{split}
\end{equation}
We also insert the identity operator which runs over all the momentum states in the propagator,
\begin{equation}
\begin{split}
\langle x_{i+1}|e^{-iH\epsilon}|x_{i}\rangle&=\int {\rm{d}}p_i \langle x_{i+1}|p_i\rangle\langle p_i|e^{-iH\epsilon}|x_{i}\rangle\\
&=\int {\rm{d}}p_i \frac{e^{ip_ix_{i+1}}}{\sqrt{2\pi}}
\langle p_i|e^{-iH\epsilon}|x_{i}\rangle .
\label{eq:app2}
\end{split}
\end{equation}
Suppose $H$ takes the form $H(x,p)=\frac{p^2}{2m}+V(x)$ and in the limit $\epsilon=\frac{t_f-t_0}{N}\to0$, we have
\begin{equation}
\exp\Big[i\big(\frac{p^2}{2m}+V(x)\big)\epsilon\Big]
\approx\exp\Big[i\frac{p^2}{2m}\epsilon\Big]\exp\Big[iV(x)\epsilon\Big],
\end{equation}
or
\begin{equation}
\langle p_i|H|x_i\rangle=\langle p_i|x_i\rangle H(x_i,p_i).
\end{equation}
The propagator in Eq.~(\ref{eq:app2}) can be written as
\begin{equation}
\begin{split}
\langle x_{i+1}|&e^{-iH\epsilon}|x_{i}\rangle=\int\frac{ {\rm{d}}p_i}{2\pi}e^{ip_i(x_{i+1}-x_i)-iH(x_i,p_i)\epsilon}\\
=&\int\frac{ {\rm{d}}p_i}{2\pi}\exp\{{i\epsilon \Big(p_i\dot{x}_{i}-\frac{p_i^2}{2m}-V(x_i)}\Big)\}\\
=&\sqrt{\frac{m}{2\pi i\epsilon}}
\exp\{i\epsilon\Big(\frac{m}{2}\dot{x}_i^2-V(x_i)\Big)\}\\
\equiv&\sqrt{\frac{m}{2\pi i \epsilon}}\exp\{i\epsilon L(x_i,\dot{x}_i)\}
\end{split}
\end{equation}
Using the above equation for each time interval in Eq.~(\ref{eq:app1}), one can get the path integral in the form, $\int \mathcal{D}x e^{iS}$, with the functional measure defined as 
\begin{equation}
\mathcal{D}x=\sqrt{\frac{m}{2\pi i\epsilon}}\prod_{i=1}^{N-1}\sqrt{\frac{m}{2\pi i\epsilon}}dx_i
\end{equation} 
Replacing $m\rightarrow \omega$, we have $A=\sqrt{\frac{2\pi i\epsilon}{\omega}}$ in \eq(\ref{K1}).

\section{Normalization of Gaussian distribution}\label{Normalization of Gaussian distribution}

The discrete version of the classical trajectory of the hard parton $\bm{R}$ in \eq(\ref{eq:R(tau)}) can be cast as
\begin{equation}
\bm{R}_j=\bm{X}+j\epsilon\frac{\bm{p}_{0}}{\omega}+\frac{\epsilon^2}{\omega}\sum_{i=1}^{j-1}(j-i)\bm{\xi}_i
\end{equation}
In this expression, $\bm{R}_j$ depends only on $\bm{\xi}_i$ with $i<j$ so that 
\begin{equation}
\int {\rm{d}}^2\bm{\xi}_j(\frac{\epsilon}{\pi})\frac{1}{\hat q(\bm{R}_j)}\exp\{-\epsilon\frac{\bm{\xi}_j^2}{\hat q(\bm{R}_j)}\}=1.
\end{equation}
Since $\hat q(\bm{R}_j)$ is a function of $\bm{\xi}_i$ with $i<j$ and does not depend on $\bm{\xi}_j$, the integrand is just a trivial Gaussian form. Then we get
\begin{equation}
\begin{split}
&\int\mathcal{D}\bm{\xi}\frac{1}{\det\{\hat q(\bm{R})\}}\exp\{-\int_{t_0}^{t_f} {\rm{d}} t\frac{\bm{\xi}^2(t)}{\hat q(\bm{R})}\}\\
=&\prod_{j=1}^{N-1}\int {\rm{d}}^2\bm{\xi}_j(\frac{\epsilon}{\pi})\frac{1}{\hat q(\bm{R}_j)}\exp\{-\epsilon\frac{\bm{\xi}_j^2}{\hat q(\bm{R}_j)}\}=1
\end{split}
\end{equation}

\section{Correlation functions}\label{On correlations}

We define the generating function
\begin{equation}
\begin{split}
Z[\bm{J}(t)]\equiv \int\mathcal{D} & \bm{\xi}(t)\frac{1}{\det\{\hat q(\bm{R})\}}
\exp\{-\int {\rm{d}}t\frac{\bm{\xi}^2(t)}{\hat q(\bm{R})}\}\\
& \times \exp\{\int {\rm{d}}t \bm{J}(t)\cdot \bm{\xi}(t)\},
\label{eq:ZJ}
\end{split}
\end{equation}
and the corresponding discretization version is
\begin{equation}
\begin{split}
Z[\bm{J}(t)]=
\prod_{j=1}^{N-1}\int {\rm{d}}^2\bm{\xi}_j(\frac{\epsilon}{\pi})& \frac{1}{\hat q(\bm{R}_j)} \exp\{-\epsilon\frac{\bm{\xi}_j^2}{\hat q(\bm{R}_j)}\}\\
\times & \exp\{-\epsilon \bm{J}_j\cdot\bm{\xi}_j\},
\end{split}
\end{equation}
which encodes all the correlation functions.

In the case of a constant $\hat q(\bm{R})={\hat q_0}$ in uniform medium,
differentiating $Z[\bm{J}(t)]$ with respect to $\bm{J}(t)$ evaluated at some time $t=t_1$ leads to
\begin{equation}
\begin{split}
\frac{\delta Z}{\delta J^a(t_1)}\Big|_{\bm{J=\bm{0}}}=&\int\mathcal{D}\bm{\xi}(t)\frac{1}{\det\{{\hat q_0}\}}\exp\{-\int {\rm{d}}t\frac{\bm{\xi}(t)^2}{{\hat q_0}}\}\xi^a(t_1)\\
=&\big\langle\xi^a(t_1)\big\rangle_0.
\end{split}
\end{equation}
In the same spirit, taking $n$ derivatives gives us,
\begin{equation}
\begin{split}
&\frac{\delta^n Z}{\delta J^{a_1}(t_1)\delta J^{a_2}(t_2)\cdots\delta J^{a_n}(t_n)}\Big|_{\bm{J=\bm{0}}}\\
=&\int\mathcal{D}\bm{\xi}(t)\frac{1}{\det\{{\hat q_0}\}}\exp\{-\int {\rm{d}}t\frac{\bm{\xi}(t)^2}{{\hat q_0}}\}\xi^{a_1}(t_1)
\cdots\xi^{a_n}(t_n)\\
=&\big\langle\xi^{a_1}(t_1)\xi^{a_2}(t_2)\cdots\xi^{a_n}(t_n)\big\rangle_0.
\end{split}
\end{equation}
The generating function above is just a functional Gaussian integral, which can be performed exactly. This leads us to
\begin{equation}
\begin{split}
Z[\bm{J}(t)]=&\int\mathcal{D}\bm{\xi}(t)\frac{1}{\det\{{\hat q_0}\}}
\exp\{-\int {\rm{d}}t\frac{1}{{\hat q_0}}(\bm{\xi}(t)-\frac{{\hat q_0}}{2}\bm{J})^2\}\\
\times& \exp\{\int {\rm{d}}t \frac{{\hat q_0}}{4}\bm{J}^2\}=\exp\{\int {\rm{d}}t \frac{{\hat q_0}}{4}\bm{J}^2\}.
\end{split}
\end{equation}
Taking derivatives with respect to $J^{a}$, we have
\begin{equation}
\begin{split}
\frac{\delta Z}{\delta J^a(t_1)}=&\frac{{\hat q_0}}{2}J^a(t_1)Z[\bm{J}]\\
\frac{\delta^2 Z}{\delta J^{a}(t_1)\delta J^{b}(t_2)}=&\frac{{\hat q_0}}{2}\delta^{ab}\delta(t_1-t_2)Z[\bm{J}]\\
&+\frac{\hat q^2_0}{4}J^{a}(t_1) J^{b}(t_2)Z[\bm{J}]
\end{split}
\end{equation}
Setting $\bm{J}=\bm{0}$, we get
\begin{equation}
\begin{split}
\langle\xi^a(t_1)\rangle_0=&\frac{\delta Z}{\delta J^a(t_1)}\Big|_{\bm{J}=\bm{0}}=0,\\
\langle\xi^a(t_1)\xi^b(t_2)\rangle_0=&\frac{\delta^2 Z}{\delta J^{a}(t_1)\delta J^{b}(t_2)}\Big|_{\bm{J}=\bm{0}}=\frac{{\hat q_0}}{2}\delta^{ab}\delta(t_1-t_2).
\end{split}
\end{equation}
Note that when $\hat{q}=\hat{q}(\bm{R})$ but not a constant ${\hat q_0}$, we do not have such simple expressions for two-point functions. Since $\bm{R}_{N-1}$ is $\bm{\xi}_{N-2}$-dependent, we can not perform the above path integral in the generating function \eq(\ref{eq:ZJ}) exactly.
%%%%%%%%%%%%%%%%%%%%%%%%%%%%%%%%%%%%%%%%%%%%%%%%%%%%%
\section{Spectra in a uniform medium}
\label{sec:Spectra in uniform medium}

%%%%%%%
\begin{widetext}
For a uniform medium, one can complete the following path integral by taking advantage of the two-point correlation function in \eq(\ref{eq:correlator})
\begin{eqnarray}
\Big\langle\exp\Big\{&i&\bm{x}\cdot\int_{t_0}^{t_f} {\rm{d}} t\bm{\xi}(t)\Big\}\Big\rangle_0
\equiv \int\widehat{\mathcal{D}}\bm{\xi} \exp\Big\{i\bm{x}\cdot\int_{t_0}^{t_f} {\rm{d}} t\bm{\xi}(t)-\int_{t_0}^{t_f} {\rm{d}} t
\frac{\bm{\xi}(t)^2}{\hat q_0}\Big\} \nonumber\\
&=& \int\widehat{\mathcal{D}}\bm{\xi} \exp\Big\{
-\int_{t_0}^{t_f} {\rm{d}} t\frac{1}{\hat q_0} 
\left[\bm{\xi}(t)-i\frac{\hat q_0}{2}\bm{x}\right]^2
-\int_{t_0}^{t_f} {\rm{d}} t \frac{\hat q_0}{4}\bm{x}^2
\Big\}=\exp\Big\{-\int_{t_0}^{t_f} {\rm{d}} t\frac{{\hat q_0}}{4} \bm{x}^2\Big\}.
\end{eqnarray}
According to \eq(\ref{p_dist}), we have
\begin{equation}
\begin{split}
\frac{d^2N}{d^2\bm{p}}
=\int {\rm{d}}^{2}\bm{X} {\rm{d}}^{2}\bm{p}_{0} &\frac{{\rm{d}}^2\bm{x}}{(2\pi)^2} \mathcal{W}(\bm{X},\bm{p}_{0};t_{0}) e^{-i\bm{x}\cdot(\bm{p}-\bm{p}_{0})} \exp\Big\{-\int_{t_0}^{t_f} {\rm{d}} t\frac{{\hat q_0}}{4}\bm{x}^2\Big\},
\end{split}
\end{equation}
which leads to \eq(\ref{eq:general_uni_p_dist}) after integrating over $\bm{x}$.
Similarly, one can get
\begin{equation}
\begin{split}
& \Big\langle\exp\Big\{i\bm{x}\cdot\int_{t_0}^{t_f} {\rm{d}} t'\int_{t_0}^{t'} {\rm{d}} t''\frac{\bm{\xi}(t'')}{t_f-t_0}\Big\}\Big\rangle_0
=\exp\Big\{-\int_{t_0}^{t_f} {\rm{d}} t\frac{{\hat q_0}}{12} \bm{x}^2\Big\}, \\
\end{split}
\end{equation}
\begin{equation}
\begin{split}
& \Big\langle\exp\Big\{i(\bm{x}-\bm{y})\cdot\int_{t_0}^{t_f} {\rm{d}} t'\int_{t_0}^{t'} {\rm{d}} t''\frac{\bm{\xi}(t'')}{t_f-t_0}+i\bm{y}\cdot\int_{t_0}^{t_f} {\rm{d}} t'\bm{\xi}(t')\Big\}\Big\rangle_0
=\exp\Big\{-\Big(\int_{t_0}^{t_f} {\rm{d}} t\frac{{\hat q_0}}{12} (\bm{x}-\bm{y})^2+\int_{t_0}^{t_f} {\rm{d}} t\frac{{\hat q_0}}{4}\bm{x}\cdot\bm{y}\Big)\Big\}, \\
\end{split}
\end{equation}

and then arrive at Eqs.~(\ref{eq:general_uni_Y_dist}) and (\ref{eq:general_uni_Wigner}) as well. 
\end{widetext}
%%%%%%%%%%%%%%%%%%%%%%%%%%%%%%%
%%
%%
%%%%%%%%%%%%%%%%%%%%%%%%%%%%%%%

\begin{widetext}
\section{Leading order (in $f$) approximation of the functional integral $F$}
\label{sec:E}

With the variable transformation $\bm{\xi}'=\bm{\xi}-i\frac{{\hat q_0}}{2}\bm{x}$,

\begin{equation}
\begin{split}
&F(\bm{x},\bm{X},\bm{p}_{0})\\
=&\exp\{-\int_{t_0}^{t_f}{\rm{d}} t\frac{\hat{q}_0}{4}\bm{x}^2\}
\int\mathcal{D}\bm{\xi}'\frac{\det\{1-f(\bm{R})\}}{\det\{\hat{q}_0\}}
\exp\Big\{\int_{t_0}^{t_f}{\rm{d}} t[\frac{\bm{\xi'}(t)^2}{\hat{q}_0}+i\bm{x}\cdot \bm{\xi'}(t)-\frac{\hat{q}_0}{4}\bm{x}^2 ]f(\bm{R})  \Big\}\exp \{-\int_{t_0}^{t_f}{\rm{d}} t\frac{\bm{\xi'}(t)^2}{\hat{q}_0} \}\\
=&\exp\{-\int_{t_0}^{t_f}{\rm{d}} t\frac{\hat{q}_0}{4}\bm{x}^2\} \prod_{k=1}^{N-1}\int\big(\frac{\epsilon}{\pi}\big)d^2\bm{\xi'}_{k}\frac{1-f(\bm{R}_k)}{\hat{q}_0}
\exp\Big\{\epsilon[\frac{{\bm{\xi'}_k}^2}{\hat{q}_0}+i\bm{x}\cdot \bm{\xi'}_k-\frac{\hat{q}_0}{4}\bm{x}^2 ]f(\bm{R}_k)  \Big\}\exp \{-\epsilon\frac{{\bm{\xi'}_k}^2}{\hat{q}_0} \},
\end{split}
\end{equation}
where $\epsilon\to0$ and $N\to\infty$. It can be approximately rewritten, to the leading order in $f$, as
\begin{equation}
\begin{split}
&F(\bm{x},\bm{X},\bm{p}_{0})\\
\approx&\exp\{-\int_{t_0}^{t_f}{\rm{d}} t\frac{\hat{q}_0}{4}\bm{x}^2\} \prod_{k=1}^{N-1}\int\big(\frac{\epsilon}{\pi}\big)d^2\bm{\xi'}_{k}\frac{1}{\hat{q}_0}
\Big\{1+\epsilon[\frac{{\bm{\xi'}_k}^2}{\hat{q}_0}+i\bm{x}\cdot \bm{\xi'}_k-\frac{\hat{q}_0}{4}\bm{x}^2 -\frac{1}{\epsilon}]f(\bm{R}_k)  \Big\}\exp \{-\epsilon\frac{{\bm{\xi'}_k}^2}{\hat{q}_0} \}\\
=&\exp\{-\int_{t_0}^{t_f}{\rm{d}} t\frac{\hat{q}_0}{4}\bm{x}^2\} 
\int\mathcal{D}\bm{\xi}'\frac{1}{\det\{\hat{q}_0\}}
\Big\{1+\int_{t_0}^{t_f} {\rm{d}} t[\frac{\bm{\xi'}(t)^2}{\hat{q}_0}+i\bm{x}\cdot \bm{\xi'}(t)-\frac{\hat{q}_0}{4}\bm{x}^2 -\frac{1}{\epsilon}]f(\bm{R})  \Big\}\exp \{-\int_{t_0}^{t_f} {\rm{d}} t\frac{\bm{\xi'}(t)^2}{\hat{q}_0} \}\\
\equiv&\exp\{-\int_{t_0}^{t_f}{\rm{d}} t\frac{\hat{q}_0}{4}\bm{x}^2\} 
\Big\{1+\int_{t_0}^{t_f} {\rm{d}} t
\Big\langle[\frac{\bm{\xi'}(t)^2}{\hat{q}_0}+i\bm{x}\cdot \bm{\xi'}(t)-\frac{\hat{q}_0}{4}\bm{x}^2 -\frac{1}{\epsilon}]f\big(\bm{R}(t,\bm{\xi}'(t))\Big\rangle_0  \Big\}
\end{split}
\end{equation}

Using the expansion of $f(\bm{R})$ (see Eq.~(\ref{eq:f_expand})) and the following correlators,
\begin{equation}
\big\langle[\bm{\xi}'(t)]^2\big\rangle_0\equiv\int\widehat{\mathcal{D}}\bm{\xi}'\exp\{-\int^{t_f}_{t_0} {\rm{d}} t\frac{\bm{\xi}'(t)^2}{\hat{q}_0}  \}\bm{\xi}'(t)^2=\hat{q}_0\delta(t-t)\\
\end{equation}
\begin{equation}
\begin{split}
\big\langle[\bm{\xi}'(t)]^2\Delta R_i(t,\bm{\xi}')\big\rangle_0
&=\int\widehat{\mathcal{D}}\bm{\xi}'\exp\{-\int^{t_f}_{t_0} {\rm{d}} t\frac{\bm{\xi}'(t)^2}{\hat{q}_0}\}\frac{1}{\omega}\int^{t}_{t_0}{\rm{d}} t_1(t-t_1)\bm{\xi}'(t)^2\xi_i'(t_1)=0\\
\end{split}
\end{equation}
\begin{equation}
\begin{split}
\big\langle[\bm{\xi}'(t)]^2\Delta R_i(t,\bm{\xi}')\Delta R_j(t,\bm{\xi}')\big\rangle_0
=&\big\langle[\bm{\xi}'(t)]^2\big\rangle_0\big\langle\Delta R_i(t,\bm{\xi}')\Delta R_j(t,\bm{\xi}')\big\rangle_0
\end{split}
\end{equation}
\begin{equation}
\begin{split}
&\big\langle[\bm{\xi}'(t)]^2\Delta R^{i_1}(t,\bm{\xi}')\Delta R^{i_2}(t,\bm{\xi}')\Delta R^{i_3}(t,\bm{\xi}')\Delta R^{i_4}(t,\bm{\xi}')\cdots\Delta R^{i_{2n-1}}(t,\bm{\xi}')\big\rangle_0 =0\\
\end{split}
\end{equation}
\begin{equation}
\begin{split}
&\big\langle[\bm{\xi}'(t)]^2\Delta R^{i_1}(t,\bm{\xi}')\Delta R^{i_2}(t,\bm{\xi}')
\cdots \Delta R^{i_{2n}}(t,\bm{\xi}')\big\rangle_0
=\big\langle[\bm{\xi}'(t)]^2\big\rangle_0\big\langle\Delta R^{i_1}(t,\bm{\xi}')\Delta R^{i_2}(t,\bm{\xi}')
\cdots\Delta R^{i_{2n}}(t,\bm{\xi}')\big\rangle_0\\
\end{split}
\end{equation}
we arrive at,
\begin{equation}
\begin{split}
F(\bm{x},\bm{X},\bm{p}_{0})
=&\exp\{ -\int_{t_0}^{t_f} {\rm{d}} t \frac{\hat{q}_0}{4} \bm{x}^2 \} \Big[1+\int_{t_0}^{t_f} {\rm{d}} t
[\frac{\hat{q}_0\delta(t-t)}{\hat{q}_0}-\frac{\hat{q}_0}{4}\bm{x}^2 -\frac{1}{\epsilon}]\Big \langle f\big(\bm{R}(t,\bm{\xi}'(t))\Big\rangle_0\Big]\\
=&\exp\{ -\int_{t_0}^{t_f} {\rm{d}} t \frac{\hat{q}_0}{4} \bm{x}^2 \} \Big[1-\int_{t_0}^{t_f} {\rm{d}} t
\frac{\hat{q}_0}{4}\bm{x}^2 \Big \langle f\big(\bm{R}(t,\bm{\xi}'(t))\Big\rangle_0 +\sum_{k=1}^{N-1} \delta_{kk}-\sum_{k=1}^{N-1}\epsilon \frac{1}{\epsilon} \Big]\\
=&\exp\{ -\int_{t_0}^{t_f} {\rm{d}} t \frac{\hat{q}_0}{4} \bm{x}^2 \} \Big[1-\int_{t_0}^{t_f} {\rm{d}} t
\frac{\hat{q}_0}{4}\bm{x}^2 \Big \langle f\big(\bm{R}(t,\bm{\xi}'(t))\Big\rangle_0 \Big]
\end{split}
\end{equation}
where we have used $\int {\rm{d}} x g(x)\delta(x-x')=g(x'\equiv x_i)=\sum_{k=1}^{N-1} g(x_k)\delta_{k i}$ for a smooth function $g(x)$.
%%%%%%%%%%%%%%%%%%%%%
\end{widetext}
%%%%%%%%%%%%%%%%%%%%%%%%%%%%%%%%%%%%%%%%%%%%%%%%%%%%%%%%%
\bibliography{main.bib}

\begin{thebibliography}{62}%
\makeatletter
\providecommand \@ifxundefined [1]{%
 \@ifx{#1\undefined}
}%
\providecommand \@ifnum [1]{%
 \ifnum #1\expandafter \@firstoftwo
 \else \expandafter \@secondoftwo
 \fi
}%
\providecommand \@ifx [1]{%
 \ifx #1\expandafter \@firstoftwo
 \else \expandafter \@secondoftwo
 \fi
}%
\providecommand \natexlab [1]{#1}%
\providecommand \enquote  [1]{``#1''}%
\providecommand \bibnamefont  [1]{#1}%
\providecommand \bibfnamefont [1]{#1}%
\providecommand \citenamefont [1]{#1}%
\providecommand \href@noop [0]{\@secondoftwo}%
\providecommand \href [0]{\begingroup \@sanitize@url \@href}%
\providecommand \@href[1]{\@@startlink{#1}\@@href}%
\providecommand \@@href[1]{\endgroup#1\@@endlink}%
\providecommand \@sanitize@url [0]{\catcode `\\12\catcode `\$12\catcode
  `\&12\catcode `\#12\catcode `\^12\catcode `\_12\catcode `\%12\relax}%
\providecommand \@@startlink[1]{}%
\providecommand \@@endlink[0]{}%
\providecommand \url  [0]{\begingroup\@sanitize@url \@url }%
\providecommand \@url [1]{\endgroup\@href {#1}{\urlprefix }}%
\providecommand \urlprefix  [0]{URL }%
\providecommand \Eprint [0]{\href }%
\providecommand \doibase [0]{http://dx.doi.org/}%
\providecommand \selectlanguage [0]{\@gobble}%
\providecommand \bibinfo  [0]{\@secondoftwo}%
\providecommand \bibfield  [0]{\@secondoftwo}%
\providecommand \translation [1]{[#1]}%
\providecommand \BibitemOpen [0]{}%
\providecommand \bibitemStop [0]{}%
\providecommand \bibitemNoStop [0]{.\EOS\space}%
\providecommand \EOS [0]{\spacefactor3000\relax}%
\providecommand \BibitemShut  [1]{\csname bibitem#1\endcsname}%
\let\auto@bib@innerbib\@empty
%</preamble>
\bibitem [{\citenamefont {Adcox}\ \emph {et~al.}(2005)\citenamefont {Adcox}
  \emph {et~al.}}]{PHENIX:2004vcz}%
  \BibitemOpen
  \bibfield  {author} {\bibinfo {author} {\bibfnamefont {K.}~\bibnamefont
  {Adcox}} \emph {et~al.} (\bibinfo {collaboration} {PHENIX}),\ }\href
  {\doibase 10.1016/j.nuclphysa.2005.03.086} {\bibfield  {journal} {\bibinfo
  {journal} {Nucl. Phys. A}\ }\textbf {\bibinfo {volume} {757}},\ \bibinfo
  {pages} {184} (\bibinfo {year} {2005})},\ \Eprint
  {http://arxiv.org/abs/nucl-ex/0410003} {arXiv:nucl-ex/0410003} \BibitemShut
  {NoStop}%
\bibitem [{\citenamefont {Arsene}\ \emph {et~al.}(2005)\citenamefont {Arsene}
  \emph {et~al.}}]{BRAHMS:2004adc}%
  \BibitemOpen
  \bibfield  {author} {\bibinfo {author} {\bibfnamefont {I.}~\bibnamefont
  {Arsene}} \emph {et~al.} (\bibinfo {collaboration} {BRAHMS}),\ }\href
  {\doibase 10.1016/j.nuclphysa.2005.02.130} {\bibfield  {journal} {\bibinfo
  {journal} {Nucl. Phys. A}\ }\textbf {\bibinfo {volume} {757}},\ \bibinfo
  {pages} {1} (\bibinfo {year} {2005})},\ \Eprint
  {http://arxiv.org/abs/nucl-ex/0410020} {arXiv:nucl-ex/0410020} \BibitemShut
  {NoStop}%
\bibitem [{\citenamefont {Back}\ \emph {et~al.}(2005)\citenamefont {Back} \emph
  {et~al.}}]{PHOBOS:2004zne}%
  \BibitemOpen
  \bibfield  {author} {\bibinfo {author} {\bibfnamefont {B.~B.}\ \bibnamefont
  {Back}} \emph {et~al.} (\bibinfo {collaboration} {PHOBOS}),\ }\href {\doibase
  10.1016/j.nuclphysa.2005.03.084} {\bibfield  {journal} {\bibinfo  {journal}
  {Nucl. Phys. A}\ }\textbf {\bibinfo {volume} {757}},\ \bibinfo {pages} {28}
  (\bibinfo {year} {2005})},\ \Eprint {http://arxiv.org/abs/nucl-ex/0410022}
  {arXiv:nucl-ex/0410022} \BibitemShut {NoStop}%
\bibitem [{\citenamefont {Adams}\ \emph {et~al.}(2005)\citenamefont {Adams}
  \emph {et~al.}}]{STAR:2005gfr}%
  \BibitemOpen
  \bibfield  {author} {\bibinfo {author} {\bibfnamefont {J.}~\bibnamefont
  {Adams}} \emph {et~al.} (\bibinfo {collaboration} {STAR}),\ }\href {\doibase
  10.1016/j.nuclphysa.2005.03.085} {\bibfield  {journal} {\bibinfo  {journal}
  {Nucl. Phys. A}\ }\textbf {\bibinfo {volume} {757}},\ \bibinfo {pages} {102}
  (\bibinfo {year} {2005})},\ \Eprint {http://arxiv.org/abs/nucl-ex/0501009}
  {arXiv:nucl-ex/0501009} \BibitemShut {NoStop}%
\bibitem [{\citenamefont {Gyulassy}\ and\ \citenamefont
  {Plumer}(1990)}]{Gyulassy:1990ye}%
  \BibitemOpen
  \bibfield  {author} {\bibinfo {author} {\bibfnamefont {M.}~\bibnamefont
  {Gyulassy}}\ and\ \bibinfo {author} {\bibfnamefont {M.}~\bibnamefont
  {Plumer}},\ }\href {\doibase 10.1016/0370-2693(90)91409-5} {\bibfield
  {journal} {\bibinfo  {journal} {Phys. Lett. B}\ }\textbf {\bibinfo {volume}
  {243}},\ \bibinfo {pages} {432} (\bibinfo {year} {1990})}\BibitemShut
  {NoStop}%
\bibitem [{\citenamefont {Wang}\ and\ \citenamefont
  {Gyulassy}(1992)}]{Wang:1992qdg}%
  \BibitemOpen
  \bibfield  {author} {\bibinfo {author} {\bibfnamefont {X.-N.}\ \bibnamefont
  {Wang}}\ and\ \bibinfo {author} {\bibfnamefont {M.}~\bibnamefont
  {Gyulassy}},\ }\href {\doibase 10.1103/PhysRevLett.68.1480} {\bibfield
  {journal} {\bibinfo  {journal} {Phys. Rev. Lett.}\ }\textbf {\bibinfo
  {volume} {68}},\ \bibinfo {pages} {1480} (\bibinfo {year}
  {1992})}\BibitemShut {NoStop}%
\bibitem [{\citenamefont {Adcox}\ \emph {et~al.}(2002)\citenamefont {Adcox}
  \emph {et~al.}}]{PHENIX:2001hpc}%
  \BibitemOpen
  \bibfield  {author} {\bibinfo {author} {\bibfnamefont {K.}~\bibnamefont
  {Adcox}} \emph {et~al.} (\bibinfo {collaboration} {PHENIX}),\ }\href
  {\doibase 10.1103/PhysRevLett.88.022301} {\bibfield  {journal} {\bibinfo
  {journal} {Phys. Rev. Lett.}\ }\textbf {\bibinfo {volume} {88}},\ \bibinfo
  {pages} {022301} (\bibinfo {year} {2002})},\ \Eprint
  {http://arxiv.org/abs/nucl-ex/0109003} {arXiv:nucl-ex/0109003} \BibitemShut
  {NoStop}%
\bibitem [{\citenamefont {Adler}\ \emph {et~al.}(2002)\citenamefont {Adler}
  \emph {et~al.}}]{STAR:2002ggv}%
  \BibitemOpen
  \bibfield  {author} {\bibinfo {author} {\bibfnamefont {C.}~\bibnamefont
  {Adler}} \emph {et~al.} (\bibinfo {collaboration} {STAR}),\ }\href {\doibase
  10.1103/PhysRevLett.89.202301} {\bibfield  {journal} {\bibinfo  {journal}
  {Phys. Rev. Lett.}\ }\textbf {\bibinfo {volume} {89}},\ \bibinfo {pages}
  {202301} (\bibinfo {year} {2002})},\ \Eprint
  {http://arxiv.org/abs/nucl-ex/0206011} {arXiv:nucl-ex/0206011} \BibitemShut
  {NoStop}%
\bibitem [{\citenamefont {Aad}\ \emph {et~al.}(2010)\citenamefont {Aad} \emph
  {et~al.}}]{ATLAS:2010isq}%
  \BibitemOpen
  \bibfield  {author} {\bibinfo {author} {\bibfnamefont {G.}~\bibnamefont
  {Aad}} \emph {et~al.} (\bibinfo {collaboration} {ATLAS}),\ }\href {\doibase
  10.1103/PhysRevLett.105.252303} {\bibfield  {journal} {\bibinfo  {journal}
  {Phys. Rev. Lett.}\ }\textbf {\bibinfo {volume} {105}},\ \bibinfo {pages}
  {252303} (\bibinfo {year} {2010})},\ \Eprint {http://arxiv.org/abs/1011.6182}
  {arXiv:1011.6182 [hep-ex]} \BibitemShut {NoStop}%
\bibitem [{\citenamefont {Chatrchyan}\ \emph {et~al.}(2011)\citenamefont
  {Chatrchyan} \emph {et~al.}}]{CMS:2011iwn}%
  \BibitemOpen
  \bibfield  {author} {\bibinfo {author} {\bibfnamefont {S.}~\bibnamefont
  {Chatrchyan}} \emph {et~al.} (\bibinfo {collaboration} {CMS}),\ }\href
  {\doibase 10.1103/PhysRevC.84.024906} {\bibfield  {journal} {\bibinfo
  {journal} {Phys. Rev. C}\ }\textbf {\bibinfo {volume} {84}},\ \bibinfo
  {pages} {024906} (\bibinfo {year} {2011})},\ \Eprint
  {http://arxiv.org/abs/1102.1957} {arXiv:1102.1957 [nucl-ex]} \BibitemShut
  {NoStop}%
\bibitem [{\citenamefont {Aamodt}\ \emph {et~al.}(2011)\citenamefont {Aamodt}
  \emph {et~al.}}]{ALICE:2010yje}%
  \BibitemOpen
  \bibfield  {author} {\bibinfo {author} {\bibfnamefont {K.}~\bibnamefont
  {Aamodt}} \emph {et~al.} (\bibinfo {collaboration} {ALICE}),\ }\href
  {\doibase 10.1016/j.physletb.2010.12.020} {\bibfield  {journal} {\bibinfo
  {journal} {Phys. Lett. B}\ }\textbf {\bibinfo {volume} {696}},\ \bibinfo
  {pages} {30} (\bibinfo {year} {2011})},\ \Eprint
  {http://arxiv.org/abs/1012.1004} {arXiv:1012.1004 [nucl-ex]} \BibitemShut
  {NoStop}%
\bibitem [{\citenamefont {Aad}\ \emph {et~al.}(2013)\citenamefont {Aad} \emph
  {et~al.}}]{ATLAS:2012tjt}%
  \BibitemOpen
  \bibfield  {author} {\bibinfo {author} {\bibfnamefont {G.}~\bibnamefont
  {Aad}} \emph {et~al.} (\bibinfo {collaboration} {ATLAS}),\ }\href {\doibase
  10.1016/j.physletb.2013.01.024} {\bibfield  {journal} {\bibinfo  {journal}
  {Phys. Lett. B}\ }\textbf {\bibinfo {volume} {719}},\ \bibinfo {pages} {220}
  (\bibinfo {year} {2013})},\ \Eprint {http://arxiv.org/abs/1208.1967}
  {arXiv:1208.1967 [hep-ex]} \BibitemShut {NoStop}%
\bibitem [{\citenamefont {Aaboud}\ \emph {et~al.}(2019)\citenamefont {Aaboud}
  \emph {et~al.}}]{ATLAS:2018gwx}%
  \BibitemOpen
  \bibfield  {author} {\bibinfo {author} {\bibfnamefont {M.}~\bibnamefont
  {Aaboud}} \emph {et~al.} (\bibinfo {collaboration} {ATLAS}),\ }\href
  {\doibase 10.1016/j.physletb.2018.10.076} {\bibfield  {journal} {\bibinfo
  {journal} {Phys. Lett. B}\ }\textbf {\bibinfo {volume} {790}},\ \bibinfo
  {pages} {108} (\bibinfo {year} {2019})},\ \Eprint
  {http://arxiv.org/abs/1805.05635} {arXiv:1805.05635 [nucl-ex]} \BibitemShut
  {NoStop}%
\bibitem [{\citenamefont {Gyulassy}\ and\ \citenamefont
  {Wang}(1994)}]{Gyulassy:1993hr}%
  \BibitemOpen
  \bibfield  {author} {\bibinfo {author} {\bibfnamefont {M.}~\bibnamefont
  {Gyulassy}}\ and\ \bibinfo {author} {\bibfnamefont {X.-n.}\ \bibnamefont
  {Wang}},\ }\href {\doibase 10.1016/0550-3213(94)90079-5} {\bibfield
  {journal} {\bibinfo  {journal} {Nucl. Phys. B}\ }\textbf {\bibinfo {volume}
  {420}},\ \bibinfo {pages} {583} (\bibinfo {year} {1994})},\ \Eprint
  {http://arxiv.org/abs/nucl-th/9306003} {arXiv:nucl-th/9306003} \BibitemShut
  {NoStop}%
\bibitem [{\citenamefont {Baier}\ \emph
  {et~al.}(1997{\natexlab{a}})\citenamefont {Baier}, \citenamefont
  {Dokshitzer}, \citenamefont {Mueller}, \citenamefont {Peigne},\ and\
  \citenamefont {Schiff}}]{Baier:1996kr}%
  \BibitemOpen
  \bibfield  {author} {\bibinfo {author} {\bibfnamefont {R.}~\bibnamefont
  {Baier}}, \bibinfo {author} {\bibfnamefont {Y.~L.}\ \bibnamefont
  {Dokshitzer}}, \bibinfo {author} {\bibfnamefont {A.~H.}\ \bibnamefont
  {Mueller}}, \bibinfo {author} {\bibfnamefont {S.}~\bibnamefont {Peigne}}, \
  and\ \bibinfo {author} {\bibfnamefont {D.}~\bibnamefont {Schiff}},\ }\href
  {\doibase 10.1016/S0550-3213(96)00553-6} {\bibfield  {journal} {\bibinfo
  {journal} {Nucl. Phys. B}\ }\textbf {\bibinfo {volume} {483}},\ \bibinfo
  {pages} {291} (\bibinfo {year} {1997}{\natexlab{a}})},\ \Eprint
  {http://arxiv.org/abs/hep-ph/9607355} {arXiv:hep-ph/9607355} \BibitemShut
  {NoStop}%
\bibitem [{\citenamefont {Baier}\ \emph
  {et~al.}(1997{\natexlab{b}})\citenamefont {Baier}, \citenamefont
  {Dokshitzer}, \citenamefont {Mueller}, \citenamefont {Peigne},\ and\
  \citenamefont {Schiff}}]{Baier:1996sk}%
  \BibitemOpen
  \bibfield  {author} {\bibinfo {author} {\bibfnamefont {R.}~\bibnamefont
  {Baier}}, \bibinfo {author} {\bibfnamefont {Y.~L.}\ \bibnamefont
  {Dokshitzer}}, \bibinfo {author} {\bibfnamefont {A.~H.}\ \bibnamefont
  {Mueller}}, \bibinfo {author} {\bibfnamefont {S.}~\bibnamefont {Peigne}}, \
  and\ \bibinfo {author} {\bibfnamefont {D.}~\bibnamefont {Schiff}},\ }\href
  {\doibase 10.1016/S0550-3213(96)00581-0} {\bibfield  {journal} {\bibinfo
  {journal} {Nucl. Phys. B}\ }\textbf {\bibinfo {volume} {484}},\ \bibinfo
  {pages} {265} (\bibinfo {year} {1997}{\natexlab{b}})},\ \Eprint
  {http://arxiv.org/abs/hep-ph/9608322} {arXiv:hep-ph/9608322} \BibitemShut
  {NoStop}%
\bibitem [{\citenamefont {Baier}\ \emph {et~al.}(1998)\citenamefont {Baier},
  \citenamefont {Dokshitzer}, \citenamefont {Mueller},\ and\ \citenamefont
  {Schiff}}]{Baier:1998kq}%
  \BibitemOpen
  \bibfield  {author} {\bibinfo {author} {\bibfnamefont {R.}~\bibnamefont
  {Baier}}, \bibinfo {author} {\bibfnamefont {Y.~L.}\ \bibnamefont
  {Dokshitzer}}, \bibinfo {author} {\bibfnamefont {A.~H.}\ \bibnamefont
  {Mueller}}, \ and\ \bibinfo {author} {\bibfnamefont {D.}~\bibnamefont
  {Schiff}},\ }\href {\doibase 10.1016/S0550-3213(98)00546-X} {\bibfield
  {journal} {\bibinfo  {journal} {Nucl. Phys. B}\ }\textbf {\bibinfo {volume}
  {531}},\ \bibinfo {pages} {403} (\bibinfo {year} {1998})},\ \Eprint
  {http://arxiv.org/abs/hep-ph/9804212} {arXiv:hep-ph/9804212} \BibitemShut
  {NoStop}%
\bibitem [{\citenamefont {Zakharov}(1996)}]{Zakharov:1996fv}%
  \BibitemOpen
  \bibfield  {author} {\bibinfo {author} {\bibfnamefont {B.~G.}\ \bibnamefont
  {Zakharov}},\ }\href {\doibase 10.1134/1.567126} {\bibfield  {journal}
  {\bibinfo  {journal} {JETP Lett.}\ }\textbf {\bibinfo {volume} {63}},\
  \bibinfo {pages} {952} (\bibinfo {year} {1996})},\ \Eprint
  {http://arxiv.org/abs/hep-ph/9607440} {arXiv:hep-ph/9607440} \BibitemShut
  {NoStop}%
\bibitem [{\citenamefont {Zakharov}(1997)}]{Zakharov:1997uu}%
  \BibitemOpen
  \bibfield  {author} {\bibinfo {author} {\bibfnamefont {B.~G.}\ \bibnamefont
  {Zakharov}},\ }\href {\doibase 10.1134/1.567389} {\bibfield  {journal}
  {\bibinfo  {journal} {JETP Lett.}\ }\textbf {\bibinfo {volume} {65}},\
  \bibinfo {pages} {615} (\bibinfo {year} {1997})},\ \Eprint
  {http://arxiv.org/abs/hep-ph/9704255} {arXiv:hep-ph/9704255} \BibitemShut
  {NoStop}%
\bibitem [{\citenamefont {Zakharov}(1998)}]{Zakharov:1998sv}%
  \BibitemOpen
  \bibfield  {author} {\bibinfo {author} {\bibfnamefont {B.~G.}\ \bibnamefont
  {Zakharov}},\ }\href@noop {} {\bibfield  {journal} {\bibinfo  {journal}
  {Phys. Atom. Nucl.}\ }\textbf {\bibinfo {volume} {61}},\ \bibinfo {pages}
  {838} (\bibinfo {year} {1998})},\ \Eprint
  {http://arxiv.org/abs/hep-ph/9807540} {arXiv:hep-ph/9807540} \BibitemShut
  {NoStop}%
\bibitem [{\citenamefont {Gyulassy}\ \emph {et~al.}(2000)\citenamefont
  {Gyulassy}, \citenamefont {Levai},\ and\ \citenamefont
  {Vitev}}]{Gyulassy:2000fs}%
  \BibitemOpen
  \bibfield  {author} {\bibinfo {author} {\bibfnamefont {M.}~\bibnamefont
  {Gyulassy}}, \bibinfo {author} {\bibfnamefont {P.}~\bibnamefont {Levai}}, \
  and\ \bibinfo {author} {\bibfnamefont {I.}~\bibnamefont {Vitev}},\ }\href
  {\doibase 10.1103/PhysRevLett.85.5535} {\bibfield  {journal} {\bibinfo
  {journal} {Phys. Rev. Lett.}\ }\textbf {\bibinfo {volume} {85}},\ \bibinfo
  {pages} {5535} (\bibinfo {year} {2000})},\ \Eprint
  {http://arxiv.org/abs/nucl-th/0005032} {arXiv:nucl-th/0005032} \BibitemShut
  {NoStop}%
\bibitem [{\citenamefont {Gyulassy}\ \emph
  {et~al.}(2001{\natexlab{a}})\citenamefont {Gyulassy}, \citenamefont {Levai},\
  and\ \citenamefont {Vitev}}]{Gyulassy:2000er}%
  \BibitemOpen
  \bibfield  {author} {\bibinfo {author} {\bibfnamefont {M.}~\bibnamefont
  {Gyulassy}}, \bibinfo {author} {\bibfnamefont {P.}~\bibnamefont {Levai}}, \
  and\ \bibinfo {author} {\bibfnamefont {I.}~\bibnamefont {Vitev}},\ }\href
  {\doibase 10.1016/S0550-3213(00)00652-0} {\bibfield  {journal} {\bibinfo
  {journal} {Nucl. Phys. B}\ }\textbf {\bibinfo {volume} {594}},\ \bibinfo
  {pages} {371} (\bibinfo {year} {2001}{\natexlab{a}})},\ \Eprint
  {http://arxiv.org/abs/nucl-th/0006010} {arXiv:nucl-th/0006010} \BibitemShut
  {NoStop}%
\bibitem [{\citenamefont {Wiedemann}(2000{\natexlab{a}})}]{Wiedemann:2000ez}%
  \BibitemOpen
  \bibfield  {author} {\bibinfo {author} {\bibfnamefont {U.~A.}\ \bibnamefont
  {Wiedemann}},\ }\href {\doibase 10.1016/S0550-3213(00)00286-8} {\bibfield
  {journal} {\bibinfo  {journal} {Nucl. Phys. B}\ }\textbf {\bibinfo {volume}
  {582}},\ \bibinfo {pages} {409} (\bibinfo {year} {2000}{\natexlab{a}})},\
  \Eprint {http://arxiv.org/abs/hep-ph/0003021} {arXiv:hep-ph/0003021}
  \BibitemShut {NoStop}%
\bibitem [{\citenamefont {Wiedemann}(2000{\natexlab{b}})}]{Wiedemann:2000za}%
  \BibitemOpen
  \bibfield  {author} {\bibinfo {author} {\bibfnamefont {U.~A.}\ \bibnamefont
  {Wiedemann}},\ }\href {\doibase 10.1016/S0550-3213(00)00457-0} {\bibfield
  {journal} {\bibinfo  {journal} {Nucl. Phys. B}\ }\textbf {\bibinfo {volume}
  {588}},\ \bibinfo {pages} {303} (\bibinfo {year} {2000}{\natexlab{b}})},\
  \Eprint {http://arxiv.org/abs/hep-ph/0005129} {arXiv:hep-ph/0005129}
  \BibitemShut {NoStop}%
\bibitem [{\citenamefont {Kovner}\ and\ \citenamefont
  {Wiedemann}(2003)}]{Kovner:2003zj}%
  \BibitemOpen
  \bibfield  {author} {\bibinfo {author} {\bibfnamefont {A.}~\bibnamefont
  {Kovner}}\ and\ \bibinfo {author} {\bibfnamefont {U.~A.}\ \bibnamefont
  {Wiedemann}},\ }\href {\doibase 10.1142/9789812795533_0004} {\ ,\ \bibinfo
  {pages} {192} (\bibinfo {year} {2003})},\ \Eprint
  {http://arxiv.org/abs/hep-ph/0304151} {arXiv:hep-ph/0304151} \BibitemShut
  {NoStop}%
\bibitem [{\citenamefont {Salgado}\ and\ \citenamefont
  {Wiedemann}(2002)}]{Salgado:2002cd}%
  \BibitemOpen
  \bibfield  {author} {\bibinfo {author} {\bibfnamefont {C.~A.}\ \bibnamefont
  {Salgado}}\ and\ \bibinfo {author} {\bibfnamefont {U.~A.}\ \bibnamefont
  {Wiedemann}},\ }\href {\doibase 10.1103/PhysRevLett.89.092303} {\bibfield
  {journal} {\bibinfo  {journal} {Phys. Rev. Lett.}\ }\textbf {\bibinfo
  {volume} {89}},\ \bibinfo {pages} {092303} (\bibinfo {year} {2002})},\
  \Eprint {http://arxiv.org/abs/hep-ph/0204221} {arXiv:hep-ph/0204221}
  \BibitemShut {NoStop}%
\bibitem [{\citenamefont {Armesto}\ \emph {et~al.}(2005)\citenamefont
  {Armesto}, \citenamefont {Salgado},\ and\ \citenamefont
  {Wiedemann}}]{Armesto:2004ud}%
  \BibitemOpen
  \bibfield  {author} {\bibinfo {author} {\bibfnamefont {N.}~\bibnamefont
  {Armesto}}, \bibinfo {author} {\bibfnamefont {C.~A.}\ \bibnamefont
  {Salgado}}, \ and\ \bibinfo {author} {\bibfnamefont {U.~A.}\ \bibnamefont
  {Wiedemann}},\ }\href {\doibase 10.1103/PhysRevLett.94.022002} {\bibfield
  {journal} {\bibinfo  {journal} {Phys. Rev. Lett.}\ }\textbf {\bibinfo
  {volume} {94}},\ \bibinfo {pages} {022002} (\bibinfo {year} {2005})},\
  \Eprint {http://arxiv.org/abs/hep-ph/0407018} {arXiv:hep-ph/0407018}
  \BibitemShut {NoStop}%
\bibitem [{\citenamefont {Arnold}\ \emph
  {et~al.}(2001{\natexlab{a}})\citenamefont {Arnold}, \citenamefont {Moore},\
  and\ \citenamefont {Yaffe}}]{Arnold:2001ba}%
  \BibitemOpen
  \bibfield  {author} {\bibinfo {author} {\bibfnamefont {P.~B.}\ \bibnamefont
  {Arnold}}, \bibinfo {author} {\bibfnamefont {G.~D.}\ \bibnamefont {Moore}}, \
  and\ \bibinfo {author} {\bibfnamefont {L.~G.}\ \bibnamefont {Yaffe}},\ }\href
  {\doibase 10.1088/1126-6708/2001/11/057} {\bibfield  {journal} {\bibinfo
  {journal} {JHEP}\ }\textbf {\bibinfo {volume} {11}},\ \bibinfo {pages} {057}
  (\bibinfo {year} {2001}{\natexlab{a}})},\ \Eprint
  {http://arxiv.org/abs/hep-ph/0109064} {arXiv:hep-ph/0109064} \BibitemShut
  {NoStop}%
\bibitem [{\citenamefont {Arnold}\ \emph
  {et~al.}(2001{\natexlab{b}})\citenamefont {Arnold}, \citenamefont {Moore},\
  and\ \citenamefont {Yaffe}}]{Arnold:2001ms}%
  \BibitemOpen
  \bibfield  {author} {\bibinfo {author} {\bibfnamefont {P.~B.}\ \bibnamefont
  {Arnold}}, \bibinfo {author} {\bibfnamefont {G.~D.}\ \bibnamefont {Moore}}, \
  and\ \bibinfo {author} {\bibfnamefont {L.~G.}\ \bibnamefont {Yaffe}},\ }\href
  {\doibase 10.1088/1126-6708/2001/12/009} {\bibfield  {journal} {\bibinfo
  {journal} {JHEP}\ }\textbf {\bibinfo {volume} {12}},\ \bibinfo {pages} {009}
  (\bibinfo {year} {2001}{\natexlab{b}})},\ \Eprint
  {http://arxiv.org/abs/hep-ph/0111107} {arXiv:hep-ph/0111107} \BibitemShut
  {NoStop}%
\bibitem [{\citenamefont {Arnold}\ \emph {et~al.}(2002)\citenamefont {Arnold},
  \citenamefont {Moore},\ and\ \citenamefont {Yaffe}}]{Arnold:2002ja}%
  \BibitemOpen
  \bibfield  {author} {\bibinfo {author} {\bibfnamefont {P.~B.}\ \bibnamefont
  {Arnold}}, \bibinfo {author} {\bibfnamefont {G.~D.}\ \bibnamefont {Moore}}, \
  and\ \bibinfo {author} {\bibfnamefont {L.~G.}\ \bibnamefont {Yaffe}},\ }\href
  {\doibase 10.1088/1126-6708/2002/06/030} {\bibfield  {journal} {\bibinfo
  {journal} {JHEP}\ }\textbf {\bibinfo {volume} {06}},\ \bibinfo {pages} {030}
  (\bibinfo {year} {2002})},\ \Eprint {http://arxiv.org/abs/hep-ph/0204343}
  {arXiv:hep-ph/0204343} \BibitemShut {NoStop}%
\bibitem [{\citenamefont {Guo}\ and\ \citenamefont {Wang}(2000)}]{Guo:2000nz}%
  \BibitemOpen
  \bibfield  {author} {\bibinfo {author} {\bibfnamefont {X.-f.}\ \bibnamefont
  {Guo}}\ and\ \bibinfo {author} {\bibfnamefont {X.-N.}\ \bibnamefont {Wang}},\
  }\href {\doibase 10.1103/PhysRevLett.85.3591} {\bibfield  {journal} {\bibinfo
   {journal} {Phys. Rev. Lett.}\ }\textbf {\bibinfo {volume} {85}},\ \bibinfo
  {pages} {3591} (\bibinfo {year} {2000})},\ \Eprint
  {http://arxiv.org/abs/hep-ph/0005044} {arXiv:hep-ph/0005044} \BibitemShut
  {NoStop}%
\bibitem [{\citenamefont {Wang}\ and\ \citenamefont
  {Guo}(2001)}]{Wang:2001ifa}%
  \BibitemOpen
  \bibfield  {author} {\bibinfo {author} {\bibfnamefont {X.-N.}\ \bibnamefont
  {Wang}}\ and\ \bibinfo {author} {\bibfnamefont {X.-f.}\ \bibnamefont {Guo}},\
  }\href {\doibase 10.1016/S0375-9474(01)01130-7} {\bibfield  {journal}
  {\bibinfo  {journal} {Nucl. Phys. A}\ }\textbf {\bibinfo {volume} {696}},\
  \bibinfo {pages} {788} (\bibinfo {year} {2001})},\ \Eprint
  {http://arxiv.org/abs/hep-ph/0102230} {arXiv:hep-ph/0102230} \BibitemShut
  {NoStop}%
\bibitem [{\citenamefont {Chen}\ \emph {et~al.}(2010)\citenamefont {Chen},
  \citenamefont {Greiner}, \citenamefont {Wang}, \citenamefont {Wang},\ and\
  \citenamefont {Xu}}]{Chen:2010te}%
  \BibitemOpen
  \bibfield  {author} {\bibinfo {author} {\bibfnamefont {X.-F.}\ \bibnamefont
  {Chen}}, \bibinfo {author} {\bibfnamefont {C.}~\bibnamefont {Greiner}},
  \bibinfo {author} {\bibfnamefont {E.}~\bibnamefont {Wang}}, \bibinfo {author}
  {\bibfnamefont {X.-N.}\ \bibnamefont {Wang}}, \ and\ \bibinfo {author}
  {\bibfnamefont {Z.}~\bibnamefont {Xu}},\ }\href {\doibase
  10.1103/PhysRevC.81.064908} {\bibfield  {journal} {\bibinfo  {journal} {Phys.
  Rev. C}\ }\textbf {\bibinfo {volume} {81}},\ \bibinfo {pages} {064908}
  (\bibinfo {year} {2010})},\ \Eprint {http://arxiv.org/abs/1002.1165}
  {arXiv:1002.1165 [nucl-th]} \BibitemShut {NoStop}%
\bibitem [{\citenamefont {Burke}\ \emph {et~al.}(2014)\citenamefont {Burke}
  \emph {et~al.}}]{JET:2013cls}%
  \BibitemOpen
  \bibfield  {author} {\bibinfo {author} {\bibfnamefont {K.~M.}\ \bibnamefont
  {Burke}} \emph {et~al.} (\bibinfo {collaboration} {JET}),\ }\href {\doibase
  10.1103/PhysRevC.90.014909} {\bibfield  {journal} {\bibinfo  {journal} {Phys.
  Rev. C}\ }\textbf {\bibinfo {volume} {90}},\ \bibinfo {pages} {014909}
  (\bibinfo {year} {2014})},\ \Eprint {http://arxiv.org/abs/1312.5003}
  {arXiv:1312.5003 [nucl-th]} \BibitemShut {NoStop}%
\bibitem [{\citenamefont {Cao}\ \emph {et~al.}(2021)\citenamefont {Cao} \emph
  {et~al.}}]{JETSCAPE:2021ehl}%
  \BibitemOpen
  \bibfield  {author} {\bibinfo {author} {\bibfnamefont {S.}~\bibnamefont
  {Cao}} \emph {et~al.} (\bibinfo {collaboration} {JETSCAPE}),\ }\href
  {\doibase 10.1103/PhysRevC.104.024905} {\bibfield  {journal} {\bibinfo
  {journal} {Phys. Rev. C}\ }\textbf {\bibinfo {volume} {104}},\ \bibinfo
  {pages} {024905} (\bibinfo {year} {2021})},\ \Eprint
  {http://arxiv.org/abs/2102.11337} {arXiv:2102.11337 [nucl-th]} \BibitemShut
  {NoStop}%
\bibitem [{\citenamefont {Miller}\ \emph {et~al.}(2007)\citenamefont {Miller},
  \citenamefont {Reygers}, \citenamefont {Sanders},\ and\ \citenamefont
  {Steinberg}}]{Miller:2007ri}%
  \BibitemOpen
  \bibfield  {author} {\bibinfo {author} {\bibfnamefont {M.~L.}\ \bibnamefont
  {Miller}}, \bibinfo {author} {\bibfnamefont {K.}~\bibnamefont {Reygers}},
  \bibinfo {author} {\bibfnamefont {S.~J.}\ \bibnamefont {Sanders}}, \ and\
  \bibinfo {author} {\bibfnamefont {P.}~\bibnamefont {Steinberg}},\ }\href
  {\doibase 10.1146/annurev.nucl.57.090506.123020} {\bibfield  {journal}
  {\bibinfo  {journal} {Ann. Rev. Nucl. Part. Sci.}\ }\textbf {\bibinfo
  {volume} {57}},\ \bibinfo {pages} {205} (\bibinfo {year} {2007})},\ \Eprint
  {http://arxiv.org/abs/nucl-ex/0701025} {arXiv:nucl-ex/0701025} \BibitemShut
  {NoStop}%
\bibitem [{\citenamefont {Wang}(2001)}]{Wang:2000fq}%
  \BibitemOpen
  \bibfield  {author} {\bibinfo {author} {\bibfnamefont {X.-N.}\ \bibnamefont
  {Wang}},\ }\href {\doibase 10.1103/PhysRevC.63.054902} {\bibfield  {journal}
  {\bibinfo  {journal} {Phys. Rev. C}\ }\textbf {\bibinfo {volume} {63}},\
  \bibinfo {pages} {054902} (\bibinfo {year} {2001})},\ \Eprint
  {http://arxiv.org/abs/nucl-th/0009019} {arXiv:nucl-th/0009019} \BibitemShut
  {NoStop}%
\bibitem [{\citenamefont {Gyulassy}\ \emph
  {et~al.}(2001{\natexlab{b}})\citenamefont {Gyulassy}, \citenamefont {Vitev},\
  and\ \citenamefont {Wang}}]{Gyulassy:2000gk}%
  \BibitemOpen
  \bibfield  {author} {\bibinfo {author} {\bibfnamefont {M.}~\bibnamefont
  {Gyulassy}}, \bibinfo {author} {\bibfnamefont {I.}~\bibnamefont {Vitev}}, \
  and\ \bibinfo {author} {\bibfnamefont {X.~N.}\ \bibnamefont {Wang}},\ }\href
  {\doibase 10.1103/PhysRevLett.86.2537} {\bibfield  {journal} {\bibinfo
  {journal} {Phys. Rev. Lett.}\ }\textbf {\bibinfo {volume} {86}},\ \bibinfo
  {pages} {2537} (\bibinfo {year} {2001}{\natexlab{b}})},\ \Eprint
  {http://arxiv.org/abs/nucl-th/0012092} {arXiv:nucl-th/0012092} \BibitemShut
  {NoStop}%
\bibitem [{\citenamefont {Gyulassy}\ \emph {et~al.}(2002)\citenamefont
  {Gyulassy}, \citenamefont {Vitev}, \citenamefont {Wang},\ and\ \citenamefont
  {Huovinen}}]{Gyulassy:2001kr}%
  \BibitemOpen
  \bibfield  {author} {\bibinfo {author} {\bibfnamefont {M.}~\bibnamefont
  {Gyulassy}}, \bibinfo {author} {\bibfnamefont {I.}~\bibnamefont {Vitev}},
  \bibinfo {author} {\bibfnamefont {X.-N.}\ \bibnamefont {Wang}}, \ and\
  \bibinfo {author} {\bibfnamefont {P.}~\bibnamefont {Huovinen}},\ }\href
  {\doibase 10.1016/S0370-2693(02)01157-7} {\bibfield  {journal} {\bibinfo
  {journal} {Phys. Lett. B}\ }\textbf {\bibinfo {volume} {526}},\ \bibinfo
  {pages} {301} (\bibinfo {year} {2002})},\ \Eprint
  {http://arxiv.org/abs/nucl-th/0109063} {arXiv:nucl-th/0109063} \BibitemShut
  {NoStop}%
\bibitem [{\citenamefont {Betz}\ \emph {et~al.}(2011)\citenamefont {Betz},
  \citenamefont {Gyulassy},\ and\ \citenamefont {Torrieri}}]{Betz:2011tu}%
  \BibitemOpen
  \bibfield  {author} {\bibinfo {author} {\bibfnamefont {B.}~\bibnamefont
  {Betz}}, \bibinfo {author} {\bibfnamefont {M.}~\bibnamefont {Gyulassy}}, \
  and\ \bibinfo {author} {\bibfnamefont {G.}~\bibnamefont {Torrieri}},\ }\href
  {\doibase 10.1103/PhysRevC.84.024913} {\bibfield  {journal} {\bibinfo
  {journal} {Phys. Rev. C}\ }\textbf {\bibinfo {volume} {84}},\ \bibinfo
  {pages} {024913} (\bibinfo {year} {2011})},\ \Eprint
  {http://arxiv.org/abs/1102.5416} {arXiv:1102.5416 [nucl-th]} \BibitemShut
  {NoStop}%
\bibitem [{\citenamefont {Zigic}\ \emph {et~al.}(2019)\citenamefont {Zigic},
  \citenamefont {Salom}, \citenamefont {Auvinen}, \citenamefont {Djordjevic},\
  and\ \citenamefont {Djordjevic}}]{Zigic:2018ovr}%
  \BibitemOpen
  \bibfield  {author} {\bibinfo {author} {\bibfnamefont {D.}~\bibnamefont
  {Zigic}}, \bibinfo {author} {\bibfnamefont {I.}~\bibnamefont {Salom}},
  \bibinfo {author} {\bibfnamefont {J.}~\bibnamefont {Auvinen}}, \bibinfo
  {author} {\bibfnamefont {M.}~\bibnamefont {Djordjevic}}, \ and\ \bibinfo
  {author} {\bibfnamefont {M.}~\bibnamefont {Djordjevic}},\ }\href {\doibase
  10.1016/j.physletb.2019.02.020} {\bibfield  {journal} {\bibinfo  {journal}
  {Phys. Lett. B}\ }\textbf {\bibinfo {volume} {791}},\ \bibinfo {pages} {236}
  (\bibinfo {year} {2019})},\ \Eprint {http://arxiv.org/abs/1805.04786}
  {arXiv:1805.04786 [nucl-th]} \BibitemShut {NoStop}%
\bibitem [{\citenamefont {Andres}\ \emph {et~al.}(2020)\citenamefont {Andres},
  \citenamefont {Armesto}, \citenamefont {Niemi}, \citenamefont {Paatelainen},\
  and\ \citenamefont {Salgado}}]{Andres:2019eus}%
  \BibitemOpen
  \bibfield  {author} {\bibinfo {author} {\bibfnamefont {C.}~\bibnamefont
  {Andres}}, \bibinfo {author} {\bibfnamefont {N.}~\bibnamefont {Armesto}},
  \bibinfo {author} {\bibfnamefont {H.}~\bibnamefont {Niemi}}, \bibinfo
  {author} {\bibfnamefont {R.}~\bibnamefont {Paatelainen}}, \ and\ \bibinfo
  {author} {\bibfnamefont {C.~A.}\ \bibnamefont {Salgado}},\ }\href {\doibase
  10.1016/j.physletb.2020.135318} {\bibfield  {journal} {\bibinfo  {journal}
  {Phys. Lett. B}\ }\textbf {\bibinfo {volume} {803}},\ \bibinfo {pages}
  {135318} (\bibinfo {year} {2020})},\ \Eprint
  {http://arxiv.org/abs/1902.03231} {arXiv:1902.03231 [hep-ph]} \BibitemShut
  {NoStop}%
\bibitem [{\citenamefont {Noronha-Hostler}\ \emph {et~al.}(2016)\citenamefont
  {Noronha-Hostler}, \citenamefont {Betz}, \citenamefont {Noronha},\ and\
  \citenamefont {Gyulassy}}]{Noronha-Hostler:2016eow}%
  \BibitemOpen
  \bibfield  {author} {\bibinfo {author} {\bibfnamefont {J.}~\bibnamefont
  {Noronha-Hostler}}, \bibinfo {author} {\bibfnamefont {B.}~\bibnamefont
  {Betz}}, \bibinfo {author} {\bibfnamefont {J.}~\bibnamefont {Noronha}}, \
  and\ \bibinfo {author} {\bibfnamefont {M.}~\bibnamefont {Gyulassy}},\ }\href
  {\doibase 10.1103/PhysRevLett.116.252301} {\bibfield  {journal} {\bibinfo
  {journal} {Phys. Rev. Lett.}\ }\textbf {\bibinfo {volume} {116}},\ \bibinfo
  {pages} {252301} (\bibinfo {year} {2016})},\ \Eprint
  {http://arxiv.org/abs/1602.03788} {arXiv:1602.03788 [nucl-th]} \BibitemShut
  {NoStop}%
\bibitem [{\citenamefont {Xu}\ \emph {et~al.}(2016)\citenamefont {Xu},
  \citenamefont {Liao},\ and\ \citenamefont {Gyulassy}}]{Xu:2015bbz}%
  \BibitemOpen
  \bibfield  {author} {\bibinfo {author} {\bibfnamefont {J.}~\bibnamefont
  {Xu}}, \bibinfo {author} {\bibfnamefont {J.}~\bibnamefont {Liao}}, \ and\
  \bibinfo {author} {\bibfnamefont {M.}~\bibnamefont {Gyulassy}},\ }\href
  {\doibase 10.1007/JHEP02(2016)169} {\bibfield  {journal} {\bibinfo  {journal}
  {JHEP}\ }\textbf {\bibinfo {volume} {02}},\ \bibinfo {pages} {169} (\bibinfo
  {year} {2016})},\ \Eprint {http://arxiv.org/abs/1508.00552} {arXiv:1508.00552
  [hep-ph]} \BibitemShut {NoStop}%
\bibitem [{\citenamefont {Shi}\ \emph {et~al.}(2019)\citenamefont {Shi},
  \citenamefont {Liao},\ and\ \citenamefont {Gyulassy}}]{Shi:2018izg}%
  \BibitemOpen
  \bibfield  {author} {\bibinfo {author} {\bibfnamefont {S.}~\bibnamefont
  {Shi}}, \bibinfo {author} {\bibfnamefont {J.}~\bibnamefont {Liao}}, \ and\
  \bibinfo {author} {\bibfnamefont {M.}~\bibnamefont {Gyulassy}},\ }\href
  {\doibase 10.1088/1674-1137/43/4/044101} {\bibfield  {journal} {\bibinfo
  {journal} {Chin. Phys. C}\ }\textbf {\bibinfo {volume} {43}},\ \bibinfo
  {pages} {044101} (\bibinfo {year} {2019})},\ \Eprint
  {http://arxiv.org/abs/1808.05461} {arXiv:1808.05461 [hep-ph]} \BibitemShut
  {NoStop}%
\bibitem [{\citenamefont {He}\ \emph {et~al.}(2022)\citenamefont {He},
  \citenamefont {Chen}, \citenamefont {Luo}, \citenamefont {Cao}, \citenamefont
  {Pang},\ and\ \citenamefont {Wang}}]{He:2022evt}%
  \BibitemOpen
  \bibfield  {author} {\bibinfo {author} {\bibfnamefont {Y.}~\bibnamefont
  {He}}, \bibinfo {author} {\bibfnamefont {W.}~\bibnamefont {Chen}}, \bibinfo
  {author} {\bibfnamefont {T.}~\bibnamefont {Luo}}, \bibinfo {author}
  {\bibfnamefont {S.}~\bibnamefont {Cao}}, \bibinfo {author} {\bibfnamefont
  {L.-G.}\ \bibnamefont {Pang}}, \ and\ \bibinfo {author} {\bibfnamefont
  {X.-N.}\ \bibnamefont {Wang}},\ }\href@noop {} {\  (\bibinfo {year}
  {2022})},\ \Eprint {http://arxiv.org/abs/2201.08408} {arXiv:2201.08408
  [hep-ph]} \BibitemShut {NoStop}%
\bibitem [{\citenamefont {Zhang}\ \emph {et~al.}(2007)\citenamefont {Zhang},
  \citenamefont {Owens}, \citenamefont {Wang},\ and\ \citenamefont
  {Wang}}]{Zhang:2007ja}%
  \BibitemOpen
  \bibfield  {author} {\bibinfo {author} {\bibfnamefont {H.}~\bibnamefont
  {Zhang}}, \bibinfo {author} {\bibfnamefont {J.~F.}\ \bibnamefont {Owens}},
  \bibinfo {author} {\bibfnamefont {E.}~\bibnamefont {Wang}}, \ and\ \bibinfo
  {author} {\bibfnamefont {X.-N.}\ \bibnamefont {Wang}},\ }\href {\doibase
  10.1103/PhysRevLett.98.212301} {\bibfield  {journal} {\bibinfo  {journal}
  {Phys. Rev. Lett.}\ }\textbf {\bibinfo {volume} {98}},\ \bibinfo {pages}
  {212301} (\bibinfo {year} {2007})},\ \Eprint
  {http://arxiv.org/abs/nucl-th/0701045} {arXiv:nucl-th/0701045} \BibitemShut
  {NoStop}%
\bibitem [{\citenamefont {Zhang}\ \emph {et~al.}(2009)\citenamefont {Zhang},
  \citenamefont {Owens}, \citenamefont {Wang},\ and\ \citenamefont
  {Wang}}]{Zhang:2009rn}%
  \BibitemOpen
  \bibfield  {author} {\bibinfo {author} {\bibfnamefont {H.}~\bibnamefont
  {Zhang}}, \bibinfo {author} {\bibfnamefont {J.~F.}\ \bibnamefont {Owens}},
  \bibinfo {author} {\bibfnamefont {E.}~\bibnamefont {Wang}}, \ and\ \bibinfo
  {author} {\bibfnamefont {X.-N.}\ \bibnamefont {Wang}},\ }\href {\doibase
  10.1103/PhysRevLett.103.032302} {\bibfield  {journal} {\bibinfo  {journal}
  {Phys. Rev. Lett.}\ }\textbf {\bibinfo {volume} {103}},\ \bibinfo {pages}
  {032302} (\bibinfo {year} {2009})},\ \Eprint {http://arxiv.org/abs/0902.4000}
  {arXiv:0902.4000 [nucl-th]} \BibitemShut {NoStop}%
\bibitem [{\citenamefont {He}\ \emph {et~al.}(2020)\citenamefont {He},
  \citenamefont {Pang},\ and\ \citenamefont {Wang}}]{He:2020iow}%
  \BibitemOpen
  \bibfield  {author} {\bibinfo {author} {\bibfnamefont {Y.}~\bibnamefont
  {He}}, \bibinfo {author} {\bibfnamefont {L.-G.}\ \bibnamefont {Pang}}, \ and\
  \bibinfo {author} {\bibfnamefont {X.-N.}\ \bibnamefont {Wang}},\ }\href
  {\doibase 10.1103/PhysRevLett.125.122301} {\bibfield  {journal} {\bibinfo
  {journal} {Phys. Rev. Lett.}\ }\textbf {\bibinfo {volume} {125}},\ \bibinfo
  {pages} {122301} (\bibinfo {year} {2020})},\ \Eprint
  {http://arxiv.org/abs/2001.08273} {arXiv:2001.08273 [hep-ph]} \BibitemShut
  {NoStop}%
\bibitem [{\citenamefont {Yang}\ \emph {et~al.}(2021)\citenamefont {Yang},
  \citenamefont {Chen}, \citenamefont {He}, \citenamefont {Ke}, \citenamefont
  {Pang},\ and\ \citenamefont {Wang}}]{Yang:2021qtl}%
  \BibitemOpen
  \bibfield  {author} {\bibinfo {author} {\bibfnamefont {Z.}~\bibnamefont
  {Yang}}, \bibinfo {author} {\bibfnamefont {W.}~\bibnamefont {Chen}}, \bibinfo
  {author} {\bibfnamefont {Y.}~\bibnamefont {He}}, \bibinfo {author}
  {\bibfnamefont {W.}~\bibnamefont {Ke}}, \bibinfo {author} {\bibfnamefont
  {L.}~\bibnamefont {Pang}}, \ and\ \bibinfo {author} {\bibfnamefont {X.-N.}\
  \bibnamefont {Wang}},\ }\href {\doibase 10.1103/PhysRevLett.127.082301}
  {\bibfield  {journal} {\bibinfo  {journal} {Phys. Rev. Lett.}\ }\textbf
  {\bibinfo {volume} {127}},\ \bibinfo {pages} {082301} (\bibinfo {year}
  {2021})},\ \Eprint {http://arxiv.org/abs/2101.05422} {arXiv:2101.05422
  [hep-ph]} \BibitemShut {NoStop}%
\bibitem [{\citenamefont {Li}\ \emph {et~al.}(2011)\citenamefont {Li},
  \citenamefont {Liu}, \citenamefont {Ma}, \citenamefont {Wang},\ and\
  \citenamefont {Zhu}}]{Li:2010ts}%
  \BibitemOpen
  \bibfield  {author} {\bibinfo {author} {\bibfnamefont {H.}~\bibnamefont
  {Li}}, \bibinfo {author} {\bibfnamefont {F.}~\bibnamefont {Liu}}, \bibinfo
  {author} {\bibfnamefont {G.-l.}\ \bibnamefont {Ma}}, \bibinfo {author}
  {\bibfnamefont {X.-N.}\ \bibnamefont {Wang}}, \ and\ \bibinfo {author}
  {\bibfnamefont {Y.}~\bibnamefont {Zhu}},\ }\href {\doibase
  10.1103/PhysRevLett.106.012301} {\bibfield  {journal} {\bibinfo  {journal}
  {Phys. Rev. Lett.}\ }\textbf {\bibinfo {volume} {106}},\ \bibinfo {pages}
  {012301} (\bibinfo {year} {2011})},\ \Eprint {http://arxiv.org/abs/1006.2893}
  {arXiv:1006.2893 [nucl-th]} \BibitemShut {NoStop}%
\bibitem [{\citenamefont {Wang}\ and\ \citenamefont
  {Zhu}(2013)}]{Wang:2013cia}%
  \BibitemOpen
  \bibfield  {author} {\bibinfo {author} {\bibfnamefont {X.-N.}\ \bibnamefont
  {Wang}}\ and\ \bibinfo {author} {\bibfnamefont {Y.}~\bibnamefont {Zhu}},\
  }\href {\doibase 10.1103/PhysRevLett.111.062301} {\bibfield  {journal}
  {\bibinfo  {journal} {Phys. Rev. Lett.}\ }\textbf {\bibinfo {volume} {111}},\
  \bibinfo {pages} {062301} (\bibinfo {year} {2013})},\ \Eprint
  {http://arxiv.org/abs/1302.5874} {arXiv:1302.5874 [hep-ph]} \BibitemShut
  {NoStop}%
\bibitem [{\citenamefont {He}\ \emph {et~al.}(2015)\citenamefont {He},
  \citenamefont {Luo}, \citenamefont {Wang},\ and\ \citenamefont
  {Zhu}}]{He:2015pra}%
  \BibitemOpen
  \bibfield  {author} {\bibinfo {author} {\bibfnamefont {Y.}~\bibnamefont
  {He}}, \bibinfo {author} {\bibfnamefont {T.}~\bibnamefont {Luo}}, \bibinfo
  {author} {\bibfnamefont {X.-N.}\ \bibnamefont {Wang}}, \ and\ \bibinfo
  {author} {\bibfnamefont {Y.}~\bibnamefont {Zhu}},\ }\href {\doibase
  10.1103/PhysRevC.91.054908} {\bibfield  {journal} {\bibinfo  {journal} {Phys.
  Rev. C}\ }\textbf {\bibinfo {volume} {91}},\ \bibinfo {pages} {054908}
  (\bibinfo {year} {2015})},\ \bibinfo {note} {[Erratum: Phys.Rev.C 97, 019902
  (2018)]},\ \Eprint {http://arxiv.org/abs/1503.03313} {arXiv:1503.03313
  [nucl-th]} \BibitemShut {NoStop}%
\bibitem [{\citenamefont {Luo}\ \emph {et~al.}(2018)\citenamefont {Luo},
  \citenamefont {Cao}, \citenamefont {He},\ and\ \citenamefont
  {Wang}}]{Luo:2018pto}%
  \BibitemOpen
  \bibfield  {author} {\bibinfo {author} {\bibfnamefont {T.}~\bibnamefont
  {Luo}}, \bibinfo {author} {\bibfnamefont {S.}~\bibnamefont {Cao}}, \bibinfo
  {author} {\bibfnamefont {Y.}~\bibnamefont {He}}, \ and\ \bibinfo {author}
  {\bibfnamefont {X.-N.}\ \bibnamefont {Wang}},\ }\href {\doibase
  10.1016/j.physletb.2018.06.025} {\bibfield  {journal} {\bibinfo  {journal}
  {Phys. Lett. B}\ }\textbf {\bibinfo {volume} {782}},\ \bibinfo {pages} {707}
  (\bibinfo {year} {2018})},\ \Eprint {http://arxiv.org/abs/1803.06785}
  {arXiv:1803.06785 [hep-ph]} \BibitemShut {NoStop}%
\bibitem [{\citenamefont {Wigner}(1932)}]{Wigner:1932eb}%
  \BibitemOpen
  \bibfield  {author} {\bibinfo {author} {\bibfnamefont {E.~P.}\ \bibnamefont
  {Wigner}},\ }\href {\doibase 10.1103/PhysRev.40.749} {\bibfield  {journal}
  {\bibinfo  {journal} {Phys. Rev.}\ }\textbf {\bibinfo {volume} {40}},\
  \bibinfo {pages} {749} (\bibinfo {year} {1932})}\BibitemShut {NoStop}%
\bibitem [{\citenamefont {Hillery}\ \emph {et~al.}(1984)\citenamefont
  {Hillery}, \citenamefont {O'Connell}, \citenamefont {Scully},\ and\
  \citenamefont {Wigner}}]{Hillery:1983ms}%
  \BibitemOpen
  \bibfield  {author} {\bibinfo {author} {\bibfnamefont {M.}~\bibnamefont
  {Hillery}}, \bibinfo {author} {\bibfnamefont {R.~F.}\ \bibnamefont
  {O'Connell}}, \bibinfo {author} {\bibfnamefont {M.~O.}\ \bibnamefont
  {Scully}}, \ and\ \bibinfo {author} {\bibfnamefont {E.~P.}\ \bibnamefont
  {Wigner}},\ }\href {\doibase 10.1016/0370-1573(84)90160-1} {\bibfield
  {journal} {\bibinfo  {journal} {Phys. Rept.}\ }\textbf {\bibinfo {volume}
  {106}},\ \bibinfo {pages} {121} (\bibinfo {year} {1984})}\BibitemShut
  {NoStop}%
\bibitem [{\citenamefont {Sadofyev}\ \emph {et~al.}(2021)\citenamefont
  {Sadofyev}, \citenamefont {Sievert},\ and\ \citenamefont
  {Vitev}}]{Sadofyev:2021ohn}%
  \BibitemOpen
  \bibfield  {author} {\bibinfo {author} {\bibfnamefont {A.~V.}\ \bibnamefont
  {Sadofyev}}, \bibinfo {author} {\bibfnamefont {M.~D.}\ \bibnamefont
  {Sievert}}, \ and\ \bibinfo {author} {\bibfnamefont {I.}~\bibnamefont
  {Vitev}},\ }\href {\doibase 10.1103/PhysRevD.104.094044} {\bibfield
  {journal} {\bibinfo  {journal} {Phys. Rev. D}\ }\textbf {\bibinfo {volume}
  {104}},\ \bibinfo {pages} {094044} (\bibinfo {year} {2021})},\ \Eprint
  {http://arxiv.org/abs/2104.09513} {arXiv:2104.09513 [hep-ph]} \BibitemShut
  {NoStop}%
\bibitem [{\citenamefont {Barata}\ \emph {et~al.}(2022)\citenamefont {Barata},
  \citenamefont {Sadofyev},\ and\ \citenamefont {Salgado}}]{Barata:2022krd}%
  \BibitemOpen
  \bibfield  {author} {\bibinfo {author} {\bibfnamefont {J.~a.}\ \bibnamefont
  {Barata}}, \bibinfo {author} {\bibfnamefont {A.~V.}\ \bibnamefont
  {Sadofyev}}, \ and\ \bibinfo {author} {\bibfnamefont {C.~A.}\ \bibnamefont
  {Salgado}},\ }\href@noop {} {\  (\bibinfo {year} {2022})},\ \Eprint
  {http://arxiv.org/abs/2202.08847} {arXiv:2202.08847 [hep-ph]} \BibitemShut
  {NoStop}%
\bibitem [{\citenamefont {Hebecker}(2000)}]{Hebecker:1999ej}%
  \BibitemOpen
  \bibfield  {author} {\bibinfo {author} {\bibfnamefont {A.}~\bibnamefont
  {Hebecker}},\ }\href {\doibase 10.1016/S0370-1573(00)00005-3} {\bibfield
  {journal} {\bibinfo  {journal} {Phys. Rept.}\ }\textbf {\bibinfo {volume}
  {331}},\ \bibinfo {pages} {1} (\bibinfo {year} {2000})},\ \Eprint
  {http://arxiv.org/abs/hep-ph/9905226} {arXiv:hep-ph/9905226} \BibitemShut
  {NoStop}%
\bibitem [{\citenamefont {Casalderrey-Solana}\ and\ \citenamefont
  {Salgado}(2007)}]{Casalderrey-Solana:2007knd}%
  \BibitemOpen
  \bibfield  {author} {\bibinfo {author} {\bibfnamefont {J.}~\bibnamefont
  {Casalderrey-Solana}}\ and\ \bibinfo {author} {\bibfnamefont {C.~A.}\
  \bibnamefont {Salgado}},\ }\href@noop {} {\bibfield  {journal} {\bibinfo
  {journal} {Acta Phys. Polon. B}\ }\textbf {\bibinfo {volume} {38}},\ \bibinfo
  {pages} {3731} (\bibinfo {year} {2007})},\ \Eprint
  {http://arxiv.org/abs/0712.3443} {arXiv:0712.3443 [hep-ph]} \BibitemShut
  {NoStop}%
\bibitem [{\citenamefont {Liang}\ \emph {et~al.}(2008)\citenamefont {Liang},
  \citenamefont {Wang},\ and\ \citenamefont {Zhou}}]{Liang:2008vz}%
  \BibitemOpen
  \bibfield  {author} {\bibinfo {author} {\bibfnamefont {Z.-t.}\ \bibnamefont
  {Liang}}, \bibinfo {author} {\bibfnamefont {X.-N.}\ \bibnamefont {Wang}}, \
  and\ \bibinfo {author} {\bibfnamefont {J.}~\bibnamefont {Zhou}},\ }\href
  {\doibase 10.1103/PhysRevD.77.125010} {\bibfield  {journal} {\bibinfo
  {journal} {Phys. Rev. D}\ }\textbf {\bibinfo {volume} {77}},\ \bibinfo
  {pages} {125010} (\bibinfo {year} {2008})},\ \Eprint
  {http://arxiv.org/abs/0801.0434} {arXiv:0801.0434 [hep-ph]} \BibitemShut
  {NoStop}%
\bibitem [{\citenamefont {Majumder}\ and\ \citenamefont
  {Muller}(2008)}]{Majumder:2007hx}%
  \BibitemOpen
  \bibfield  {author} {\bibinfo {author} {\bibfnamefont {A.}~\bibnamefont
  {Majumder}}\ and\ \bibinfo {author} {\bibfnamefont {B.}~\bibnamefont
  {Muller}},\ }\href {\doibase 10.1103/PhysRevC.77.054903} {\bibfield
  {journal} {\bibinfo  {journal} {Phys. Rev. C}\ }\textbf {\bibinfo {volume}
  {77}},\ \bibinfo {pages} {054903} (\bibinfo {year} {2008})},\ \Eprint
  {http://arxiv.org/abs/0705.1147} {arXiv:0705.1147 [nucl-th]} \BibitemShut
  {NoStop}%
\end{thebibliography}%
\end{document}